\documentclass{emulateapj}
\usepackage{apjfonts}
\usepackage{natbib,amsmath}

\newcommand{\chandra}{{\it Chandra\/}}
\def\asca{{\it ASCA\/}}

\newcommand{\lum}{{erg~s$^{-1}$}}

\begin{document}

\title{DEEP CHANDRA MONITORING OBSERVATIONS OF NGC 4649: I. CATALOG OF SOURCE PROPERTIES}

\author{
B.~Luo,\altaffilmark{1,2}
G.~Fabbiano,\altaffilmark{1}
J.~Strader,\altaffilmark{1}
D.-W.~Kim,\altaffilmark{1}
J.~P.~Brodie,\altaffilmark{3}
T.~Fragos,\altaffilmark{1}
J.~S.~Gallagher,\altaffilmark{4}
A.~King,\altaffilmark{5}
and A.~Zezas\altaffilmark{6}
}
\altaffiltext{1}{
Harvard-Smithsonian Center for Astrophysics,
60 Garden Street, Cambridge, MA 02138, USA}
\altaffiltext{2}{
Department of Astronomy \& Astrophysics, 525 Davey Lab,
The Pennsylvania State University, University Park, PA 16802, USA}
\altaffiltext{3}{
UCO/Lick Observatory, 1156 High St., Santa Cruz, CA
95064, USA}
\altaffiltext{4}{
Department of Astronomy, University of Wisconsin, Madison,
WI 53706-1582, USA}
\altaffiltext{5}{
Department of Physics \& Astronomy, University of Leicester,
University Road, Leicester LE1 7RH, UK}
\altaffiltext{6}{
Physics Department, University of Crete, P.O. Box 2208,
GR-710 03, Heraklion, Crete, Greece}

\begin{abstract}
We present the X-ray source catalog for the \chandra\ monitoring observations of 
the elliptical galaxy, NGC 4649. 
The galaxy has been observed with \chandra\ ACIS-S3
in six separate pointings, reaching a total exposure of 299 ks.
There are 501 X-ray sources detected in the 0.3--8.0 keV band in the merged
observation or in one of the six individual observations; 399 sources are located 
within the D$_{25}$ ellipse. 
The observed 0.3--8.0 keV
luminosities of these 501 sources range from
$9.3\times10^{36}$~\lum\ to $5.4\times10^{39}$~\lum. The 90\% detection 
completeness 
limit within the D$_{25}$ ellipse is $5.5\times10^{37}$~\lum.
Based on the surface density of background active galactic nuclei (AGNs) and
detection completeness, we expect
$\approx45$
background AGNs
among the catalog sources ($\approx15$ within the D$_{25}$ ellipse).
There are nine sources with luminosities greater than $10^{39}$~\lum, which are candidates for
ultraluminous X-ray sources.
The nuclear source of NGC 4649 is a low-luminosity AGN, 
with an intrinsic 2.0--8.0
keV X-ray luminosity of $1.5\times10^{38}$~\lum. 
The X-ray colors suggest that the majority of the catalog sources are low-mass X-ray binaries (LMXBs).
We find that 164 of the 501 X-ray sources show long-term variability, 
indicating that they are 
accreting compact objects. We discover four transient candidates and another four
potential transients. 
We also identify 173 X-ray sources (141 within the D$_{25}$ ellipse) 
that are associated with globular clusters (GCs) based on {\it Hubble Space Telescope} and ground-based data; these LMXBs tend to be hosted by red GCs.
Although NGC 4649 has a much larger population of X-ray sources 
than the structurally similar early-type galaxies, NGC 3379 and NGC 4278, 
yet the X-ray source properties are comparable in all three systems. 
\end{abstract}

\section{INTRODUCTION}

Low-mass X-ray binaries (LMXBs) are 
binaries composed of an accreting neutron star
or black hole and a low-mass late-type companion star. 
As a trace fossil of the old stellar populations in early-type
galaxies, the origin and evolution of LMXBs have received
much attention since they were first discovered in the Milky Way 
\citep[e.g.,][]{Giacconi1974}.
It has been found that a significant fraction (20\%--70\%) of 
LMXBs are residing in globular clusters 
(GCs; e.g., \citealt{Sarazin2000,Angelini2001,Blanton2001,Kundu2002,Kim2006}), 
suggesting that
GCs play an important or even exclusive role in the formation of LMXBs 
\citep[e.g,][]{Verbunt2006,Kundu2007,Humphrey2008}. 

With the subarcsecond angular resolution of \chandra,
we are now able to reveal the X-ray binary (XRB) populations
of distant ($\approx20$--30 Mpc) galaxies.
X-ray color--color diagrams and X-ray luminosity functions (XLFs)
have been used to probe the different XRB populations, e.g., 
LMXBs that are associated with old stellar populations and high-mass
X-ray binaries (HMXBs) that are associated with young stellar
populations (see \citealt{Fabbiano2006} for a review).
\chandra\ observations have greatly extended the LMXB samples and 
improved our understanding of 
the formation and evolution of LMXBs 
and the role of GCs in these processes.
Deep \chandra\ monitoring observations have also detected 
LMXB populations down to limiting luminosity of a few $10^{36}$~\lum,
well within the luminosity range of Galactic LMXBs \citep[e..g,][]{Brassington2008,Brassington2009}.

NGC 4649 (M60) is a giant Virgo elliptical galaxy at 
a distance of $\approx17$ Mpc.
It has a companion spiral galaxy, NGC 4647, that is $2.6\arcmin$ away
in projection.
Independent distance measurements indicate that the two galaxies are 
physically close to each other and are likely gravitationally interacting 
\citep[e.g.,][and references therein]{Young2006}.
Early-type galaxies are ideal targets for constructing relatively clean
samples of LMXBs and studying the GC-LMXB association, as 
they have little contamination from the young HMXB populations and are in general
abundant in GCs \citep[e.g.,][]{Ashman1998}.
NGC 4649 has a rich GC system \citep{Harris1991}, and 
earlier studies have shown that its X-ray source population
is large, with 165 sources
detected in a $\approx20$~ks \chandra\ observation \citep{Randall2004}.
It has a remarkably large number of sources with $L_{\rm X}>2\times10^{38}$~\lum, thus likely black-hole binaries; such large populations of luminous
X-ray sources are rarely seen in elliptical galaxies.
As part of a continuing effort to obtain deep \chandra\ LMXB samples and to probe 
their formation and evolution,
we acquired an additional $\approx200$~ks \chandra\ exposure of 
NGC 4649 in the year 2011, making a total exposure of $\approx300$~ks.
Combined with our previous deep observations of the early-type galaxies 
NGC 3379 \citep{Brassington2008}
and NGC 4278 \citep{Brassington2009}, these data
provide unprecedented LMXB samples for constraining the nature of these
XRB populations \citep[e.g.,][]{Fragos2008,Kim2009}.
Multiepoch observations of NGC 4649 spanning $>10$ years also
allow variability studies and reveal the X-ray transient population that could 
pose crucial constraints to the processes and evolution of accretion disks in LMXBs \citep[e.g.,][]{Fragos2009}.

In this paper, we present a detailed \chandra\ source catalog for NGC 4649 along with
analyses of X-ray properties and source variabilities.
Detailed subsequent investigations and scientific
interpretation of the X-ray source sample 
will be presented in future
papers, e.g., studies of the XLFs (D.-W.~Kim et al. 2012, in prep.),
the GC population \citep{Strader2012}, and ultraluminous
X-ray sources (ULXs; \citealt{Roberts2012}).
In Section 2 we describe the observations and data reduction. In Section 3 
we present the source catalog and describe the method used to create this catalogs.
We discuss basic X-ray properties of the detected sources, 
including the radial profile of the LMXB surface density, 
the nuclear source, hardness ratios (HRs), X-ray colors,
and variabilities. We also present optical identifications and
GC-LMXB associations for this galaxy. We summarize in Section 4.

We adopt a distance of 16.5 Mpc to NGC 4649 \citep{Blakeslee2009}.
The nuclear position of the galaxy is $\alpha_{\rm J2000.0}=
12^{\rm h}43^{\rm m}39.97^{\rm s}$ and $\delta_{\rm J2000.0}=11\degr33\arcmin09.7\arcsec$
from the Sloan Digital Sky Survey
Data Release 7 \citep{Abazajian2009}.
The Galactic column density along the line of sight to
NGC 4649 is \hbox{$N_{\rm H}=2.2\times 10^{20}$~cm$^{-2}$} \citep{Dickey1990}.

\section{OBSERVATIONS AND DATA REDUCTION}

NGC 4649 has been covered by six \chandra\ observations
with the S3 chip of the Advanced CCD Imaging
Spectrometer (ACIS; \citealt{Garmire2003}), spanning 11
years. Table~\ref{tbl-obs} lists the six observations
along with their exposure times,
ranging from 14~ks to 102~ks.
We reduced and analyzed the observational data 
using mainly the \chandra\ Interactive Analysis
of Observations (CIAO) tools. \footnote{See
http://cxc.harvard.edu/ciao/ for details on CIAO.}
We used the {\sc chandra\_repro} script to reprocess the
data with the latest calibration.
The background light curve of each observation was then inspected
and background flares were removed using the {\sc deflare}
CIAO script, which performed an iterative 3$\sigma$ clipping algorithm. 
The flare-cleaned exposure times are also listed in Table~\ref{tbl-obs}; the total usable exposure is 299.4 ks.

We registered the astrometric
frames of all the observations to that of observation
12976, which has the longest exposure.
We created a 0.3--8.0 keV image for each observation and searched for sources
using {\sc wavdetect} \citep{Freeman2002}
at a false-positive probability threshold of 10$^{-6}$. 
Using the CIAO script {\sc reproject\_aspect}, we compared 
the source list of each individual observation to the source list
of observation
12976, adopting a 3$\arcsec$ matching radius and a
residual rejection limit of 0.6$\arcsec$, and then registered the 
astrometric
frame of the given observation to that of 
observation
12976.
We reprojected the registered observations to the frame of observation
12976 using {\sc reproject\_events}, and merged all the observations
to create a master event file using {\sc dmmerge}.
The ACIS-S3 chip has different pointings for the six observations, and
the average aim point (weighted by exposure time) is
$\alpha_{\rm aim, J2000.0}=
12^{\rm h}43^{\rm m}39.88^{\rm s}$, $\delta_{\rm aim, J2000.0}=11\degr33\arcmin06.3\arcsec$.

We created images from the merged event file using the standard
\asca\ grade set (\asca\ grades 0, 2, 3, 4, 6) for five bands (also listed
in Table~\ref{tbl-bands}):
\hbox{0.3--8.0~keV} (full band; FB), \hbox{0.3--2.0~keV} (soft band; SB),
\hbox{2.0--8.0~keV} (hard band; HB), \hbox{0.3--1.0~keV} (soft band 1; SB1),
and \hbox{1.0--2.0~keV} (soft band 2; SB2).
For each observation, we created exposure maps in these bands
following the basic procedure
outlined in Section 3.2 of \citet{Hornschemeier2001}, which
takes into account the effects of vignetting, gaps between the CCDs,
bad-column filtering, bad-pixel filtering, and the spatially
dependent degradation in quantum efficiency
due to contamination on the ACIS optical-blocking filters.
A photon index of
$\Gamma=1.7$ was assumed in creating the exposure map,
which is a typical value for XRBs
\citep[e.g.,][]{Irwin2003,Brassington2010}. 
Merged exposure maps were then created from
the exposure maps of the individual observations.

\begin{figure}
\centerline{
\includegraphics[scale=0.5]{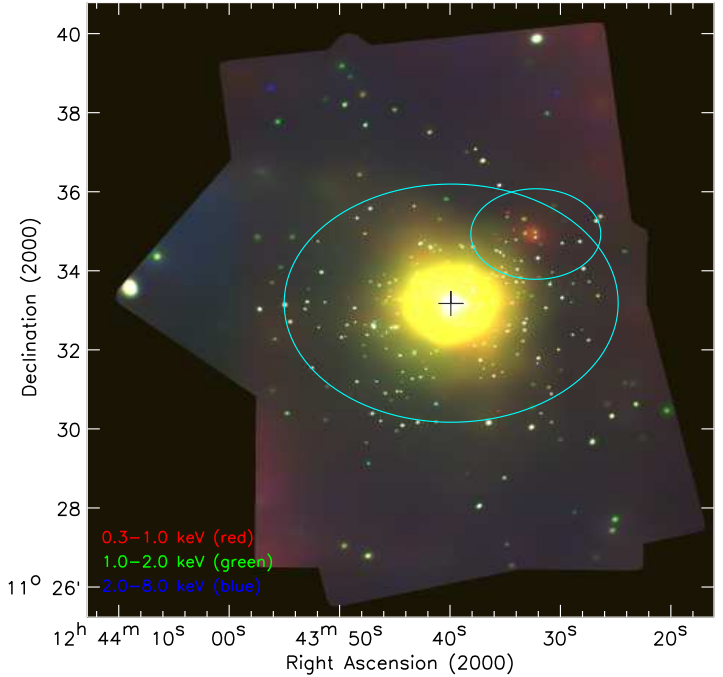}
}
\figcaption{
\chandra\ ``false-color'' image of NGC 4649. This image
is a color composite of the  exposure-corrected adaptively smoothed images
in the SB1 (red), SB2 (green), and HB (blue).
The cross symbol shows the center of the galaxy.
The cyan ellipses outline the D$_{25}$ regions 
of NGC 4649 (larger one) and the companion galaxy NGC 4647 (smaller one), 
respectively
\citep{deVaucouleurs1991}.
\label{clrimg}}
\end{figure}

We constructed adaptively smoothed images from the raw images using the CIAO
tool {\sc csmooth}.
Exposure-corrected smoothed images were then constructed following
Section 3.3 of \citet{Baganoff2003}.
We show in
Figure~\ref{clrimg} a color composite of the exposure-corrected
smoothed images
in the SB1, SB2, and HB.

\section{X-RAY SOURCE CATALOG}
\subsection{Source Detection and Photometry Extraction}
X-ray sources in NGC 4649 were searched for in the merged FB
image as well as the FB images for the six individual observations.
We adopted a two-step source-detection approach.
We first generated a candidate source list using
{\sc wavdetect};
then we 
utilized the {\sc acis extract} (AE; \citealt{Broos2010}) program
to remove low-significance sources 
(with AE binomial no-source probabilities $P_{\rm B}>0.01$) 
from the candidate list, and composed the final catalog with 
the remaining sources.
Such a two-step approach has been employed to create reliable 
\chandra\ source catalogs \citep[e.g.,][]{Broos2007,Broos2011,Xue2011}. 

{\sc wavdetect} was run
with a ``$\sqrt{2}$~sequence'' of wavelet scales (i.e.,\ 1, $\sqrt{2}$, 2,
$2\sqrt{2}$, 4, $4\sqrt{2}$, 8, $8\sqrt{2}$, and 16 pixels)
and a false-positive probability threshold of 10$^{-6}$. 
There were 471 X-ray sources detected in the merged observation,
and 88--263 sources detected in the six individual observations.
We merged the seven source lists using a matching radius 
of $2\arcsec$ for sources
within $6\arcmin$ of the average aim point and $3\arcsec$ for
larger off-axis angles; for a source detected in multiple observations
(including the merged one),
we adopted its position from the observation with the longest exposure.
The resulting candidate source list contains 517 X-ray sources.
Given the 10$^{-6}$ {\sc wavdetect} threshold, 
we expect approximately eight false detections in
the candidate list with this {\sc wavdetect}
approach (two from the merged observation 
and one from each of the six individual observations; \citealt{Kim2004}).

The photometry of the 517 candidate sources was extracted by AE.
For each source in a given observation, AE constructed a polygonal 
source-count
extraction region that approximates the
$\approx90\%$ encircled-energy fraction (EEF) contour of the point 
spread function (PSF)
at 1.5 keV. Smaller extraction regions (as low as $\approx40\%$ EEF)
were used for sources in crowded regions to avoid source overlapping.
Background counts were extracted by AE
in an annular region around each source
excluding the overlapping areas that belong to neighboring sources.
The extracted source counts and background counts were summed up over all
the observations, and the expected number of background counts in 
the source region was then 
calculated considering the scaling of the areas of the background
and source regions. A background scaling factor of $\approx6$--30 was
generally chosen by AE for our sources. The total numbers of 
extracted background counts in the merge observation range from $\approx80$
to 250 for sources not in the galactic center 
(off-axis angle $>0.5\arcmin$),
and toward the center the number increases rapidly,  
reaching a few thousand background counts for the innermost sources.
Given the extract sources counts, background counts, and background scaling
of a source, 
AE computed a binomial probability ($P_{\rm B}$) of 
observing the source counts by chance 
under the assumption that there is no source at the location (all the 
observed counts are background). 
A larger value of $P_{\rm B}$ indicates that the source has a larger chance
of being a spurious detection.
We adopted a threshold of $P_{\rm B}\le0.01$
to select reliable X-ray sources; 16 sources are removed this way (note that 
$\approx8$ false detections are expected from the {\sc wavdetect} approach). 
This
threshold value was chosen to balance the goals of removing most of the 
spurious sources and of not missing too many real sources.
The final catalog includes 501 X-ray sources. 
We note that some
of the sources are not covered by all the six observations due
to the different pointings and roll angles of the observations.

For each of the 501 sources, we derived its 
aperture-corrected net (background-subtracted) counts in the 
five bands (FB, SB, HB, SB1, SB2)
based on the AE extraction results.
AE provided EEFs at five energies (ranging from 0.28 to 8.60 keV) 
with the given extraction region, and thus we obtained the 
aperture correction for every band via interpolation.
Source counts were computed in the merged observation as well as the 
six individual observations. For a given band and a given observation
(including the merged one), we consider a source
being detected if its $P_{\rm B}$ value is smaller than 0.01 
in this band and this observation; 
otherwise the source is flagged as undetected. 
All the 501 sources must have been detected 
in the FB in at least one of the observations based on 
our source-detection approach above.
For detected sources, we adopted their AE-generated  
1$\sigma$ errors \citep{Gehrels1986} for the net counts, which
were propagated through the errors of the extracted source and
background counts following the numerical
method described in Section 1.7.3 of \citet{Lyons1991}.
For undetected sources, 3$\sigma$ upper limits on the net counts were
calculated. If the extracted number of source counts is less than 10, we 
derived the upper limit using the Bayesian approach of \citet{Kraft1991} for the 99.87\% ($\approx3\sigma$) confidence level; otherwise,
we calculated the 3$\sigma$ upper limit following the Poisson
statistic \citep{Gehrels1986}.
In the merged observation, the number of FB net counts has a range
of $\approx8$--3870.

We estimated the source positional uncertainties 
using the empirical relation proposed 
by \citet{Kim2007}. The positional uncertainty
at the 95\% confidence level is given by
\begin{equation} {\log {\rm PU}} = \kern-4pt\left\lbrace\!\! \begin{array} {r@{\quad \quad }l} 0.1145\times \mathrm{OAA}- 0.4958 \times {\log \mathrm{NC}} +0.1932, \\ \quad 0.0000 < {\log \mathrm{NC}} \le 2.1393\\ \!\!0.0968\times \mathrm{OAA} - 0.2064 \times {\log \mathrm{NC}} - 0.4260, \\ \quad 2.1393 < {\log \mathrm{NC}} \le 3.3000 \end{array}\right., \end{equation}
where PU is the positional uncertainty in arcseconds, OAA is the off-axis angle 
in arcminutes, and NC is the number of FB source counts extracted by {\sc wavdetect}.

X-ray flux and luminosity in the FB were calculated for each source in each
observation, utilizing the photometric and 
spectral information extracted by AE.
For a relatively bright source (FB net counts $\ge50$),
we fit the source spectrum using XSPEC 
(version 12.7.0; \citealt{Arnaud1996}), employing an
absorbed power-law model ({\sc tbabs*pow}) with the
Cash fitting statistic (CSTAT; \citealt{Cash1979}).
Both the photon index ($\Gamma$) and absorption were set as free parameters.
The observed FB flux was then obtained from
the best-fit model.

For the less luminous sources (FB net counts $<50$), we did not adopt
their fluxes from the spectral fitting. Instead, we converted their 
FB count rates to the FB fluxes using a count-rate-to-flux conversion
relation that is a function of the source photon index
\citep[e.g.,][]{Luo2008,Xue2011}; this conversion relation was calibrated
using the count rates, fluxes, and photon indices of those more 
luminous sources above (FB net counts $\ge50$). For a source that
is faint (FB net counts $\le30$), we assumed $\Gamma=1.7$ to convert
its count rate to flux. For the other sources (FB net counts in the range 
of 30--50), we still adopted their photon indices from the same
absorbed power-law spectral fitting as above (with $\Gamma$ and absorption
set free), and then converted their count rates to fluxes.
In the two shortest observations, 8507 and 14328, there are not enough
bright sources to constrain the count-rate-to-flux conversion, and thus
we used the relation derived for the merged observation.
For undetected sources, 3$\sigma$ upper limits on the fluxes
were converted from the count-rate upper limits assuming $\Gamma=1.7$.

FB luminosities were calculated given the FB fluxes and the 
distance of NGC 4649.
The 1$\sigma$ errors of the fluxes and luminosities were
propagated through the errors of the net counts.\footnote{We 
did not consider the flux or luminosity errors contributed by 
the uncertainties of the photon indices and/or the 
count-rate-to-flux conversion, which may be of the same order of
magnitude as the errors propagated through the counts errors. 
Detailed spectral analyses are
needed to properly determine the flux or luminosity 
errors, which are only possible for sources with sufficient counts.
The flux or luminosities errors are not used in any scientific analyses in
this work, and are only shown in some plots for illustration purposes.}
The luminosities have been corrected for Galactic absorption 
assuming $\Gamma=1.7$ (a factor of 1.055 for the FB); the corrections do
not depend significantly on the photon indices and are accurate to 
a few percent. We did not correct for any intrinsic absorption if present;
detailed spectral analyses are required to accurately determine 
the intrinsic luminosities.
In the merged observation, the FB luminosity ranges from 
$9.3\times10^{36}$~\lum\ to $5.4\times10^{39}$~\lum.

\begin{figure}
\centerline{\includegraphics[scale=0.5]{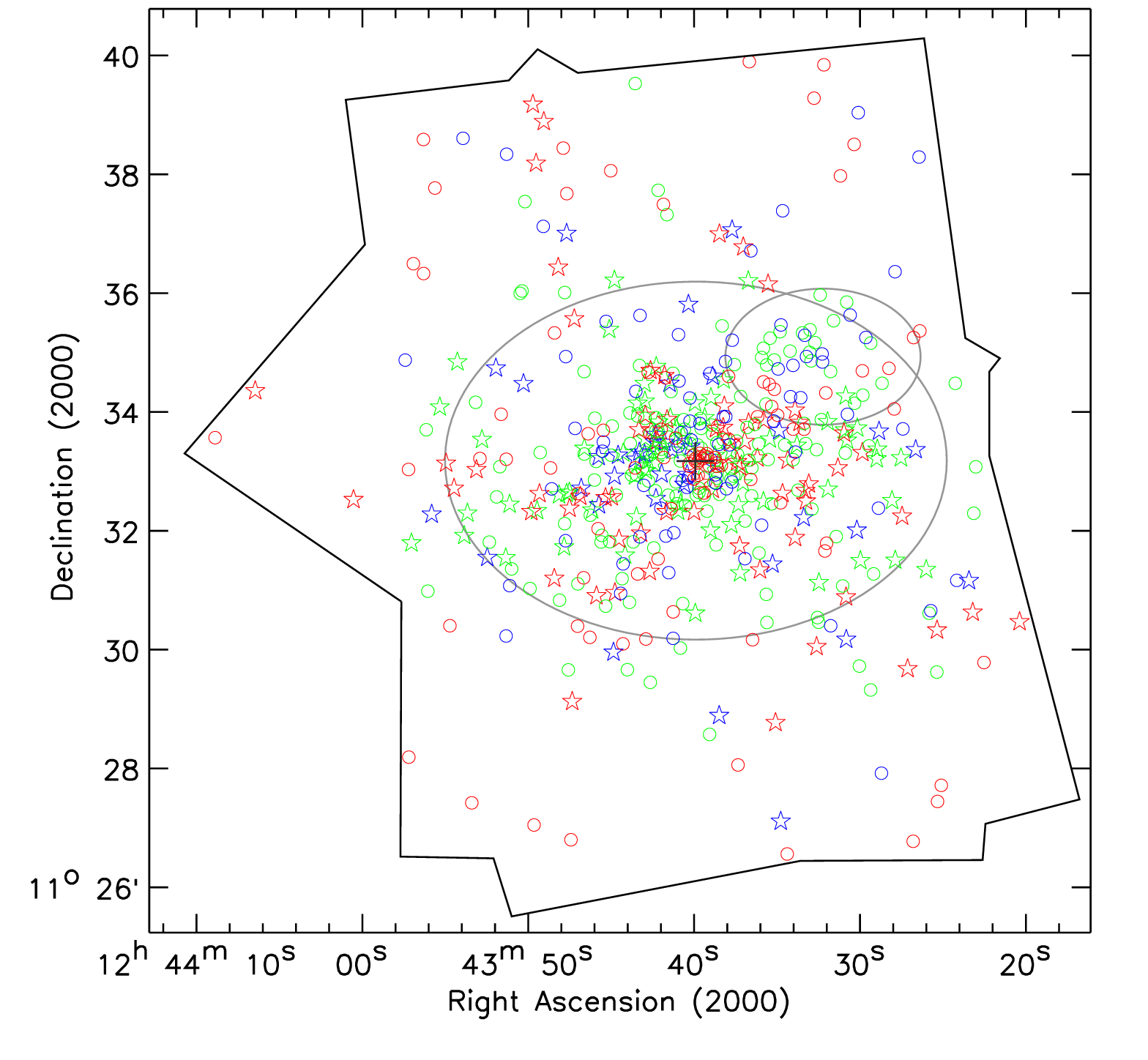}
}
\figcaption{
Positions of the 501 X-ray sources in different
luminosity bins: $>10^{38}$~\lum\ (red),
$5\times10^{37}\le L_{\rm 0.3\textrm{--}8~keV}<10^{38}$~\lum\ (blue),
and $<5\times10^{37}$~\lum\ (green).
Star symbols represents sources with a GC counterpart
(see Section~\ref{sec-gccp}), and
circles are the other sources.
The outer region shows the \chandra\ field of view.
The grey ellipses outline the D$_{25}$ regions
of NGC 4649 and NGC 4647.
(A color version of this figure is available in the online journal.)
\label{fig-pos}}
\end{figure}

The positions of the 501 sources are indicated in Figure~\ref{fig-pos},
divided into different luminosity bins. Within the \chandra\ field of view
(D$_{25}$ ellipse of NGC 4649), there are 32\% (29\%), 22\% (22\%), and 
46\% (49\%) X-ray sources with FB luminosity $>10^{38}$~\lum,
$5\times10^{37}\le L_{\rm 0.3\textrm{--}8~keV}<10^{38}$~\lum, 
and $<5\times10^{37}$~\lum, respectively.

\subsection{Source Detection Completeness} \label{completeness}
The source detection completeness of the catalog varies
across the field, mainly due to the different pointings and 
roll angles of the observations, the degradation of \chandra\
sensitivity at large off-axis angles, and the elevated background level
near the galactic center originated from diffused gas emission.
We performed simulations to assess the completeness of the catalog
following the procedures described in \citet{Kim2004b}, taking into account
the effects of flux detection limit and source confusion.
In each simulation we added a mock X-ray sources at a random location
on the event file of every observation using the MARX ray-tracing
simulator,\footnote{See http://space.mit.edu/CXC/MARX/index.html.}
and then we applied the same image creation and source detection
method as illustrated above to determine if this additional 
source is detectable.
The input
X-ray luminosity of the source was randomly drawn from a power-law XLF with
$\beta=1$ in a cumulative form [$N(>L_{\rm X})=kL_{\rm X}^{-\beta}$], and
we assumed a power-law spectrum for the source with $\Gamma=1.7$.
The position of the source was randomly selected following the $r^{1/4}$ law
\citep{deVaucouleurs1948}. We note that the adopted luminosity and position
distributions here do not affect the completeness estimation significantly,
as we only aimed to derive the positional-dependent
detection fractions at a given luminosity. During the source filtering process
($P_{\rm B}\le0.01$), instead of using AE to extract the source photometry,
we adopted the {\sc wavdetect} source and background counts, and assumed a 
typical background scaling factor of 16. This simplification greatly reduced the 
computation
time and does not affect the simulation results significantly, as the 
chance of removing a detection is small during this step (16/517 for our 
source catalog above).

\begin{figure}
\centerline{\includegraphics[scale=0.5]{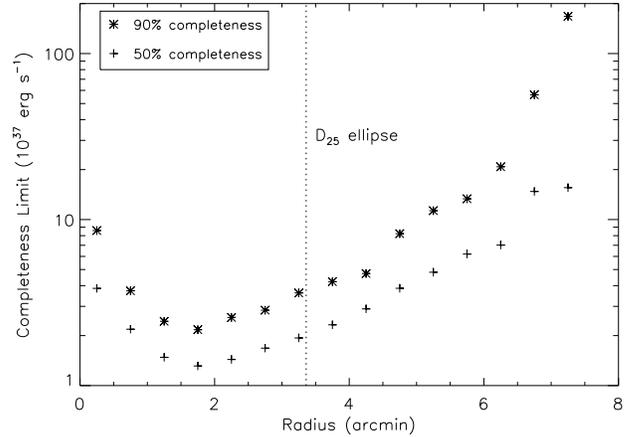}
}
\figcaption{
The 50\% and 90\% detection completeness limits as a function
of the radius to the galactic nucleus.
The completeness limits were computed in annular regions with a bin size
of 0.5\arcmin; the central 10\arcsec-radius area was excluded 
as the completeness calculations are not reliable in
this crowded region.
The vertical
dotted line indicates the average radial distance of the NGC 4649
D$_{25}$ ellipse. 
The combined \chandra\ observations are most sensitive around a 
radius of $\approx1.8\arcmin$;
at larger radii, the sensitive drops due to the lower effective exposure,
and at smaller radii, the sensitive also drops because of the strong 
background level coming from diffused gas emission in the galactic center.
\label{fig-comp}}
\end{figure}

We performed 90,000 simulations in total, 
and we computed the probability of detecting a source with 
a given luminosity in a given region, utilizing the properties of all the
simulated sources 
in this region. The 50\% or 90\% 
detection completeness limit in this region was then derived via interpolation.
In Figure~\ref{fig-comp}, we show the 50\% and 90\% completeness limits
as a function of the galactic radius.
The central 10\arcsec-radius area was excluded from the calculation
as the completeness estimation is not reliable in
this crowded region.
The highest sensitivity is reached at a radius of $\approx1.8\arcmin$,
with a 50\% (90\%) completeness limit of $1.3\times10^{37}$ 
($2.2\times10^{37}$)~\lum. At larger radii, the sensitive 
drops due to the lower effective exposure;
at smaller radii, the sensitive also drops because of the strong 
background level coming from diffused gas emission in the galactic center.
The average 50\% (90\%) completeness limit of the D$_{25}$ region
is $2.1\times10^{37}$ ($5.5\times10^{37}$)~\lum.

\subsection{Radial Profile of X-ray Sources} \label{radprofile}
\begin{figure*}
\centerline{
\includegraphics[scale=0.5]{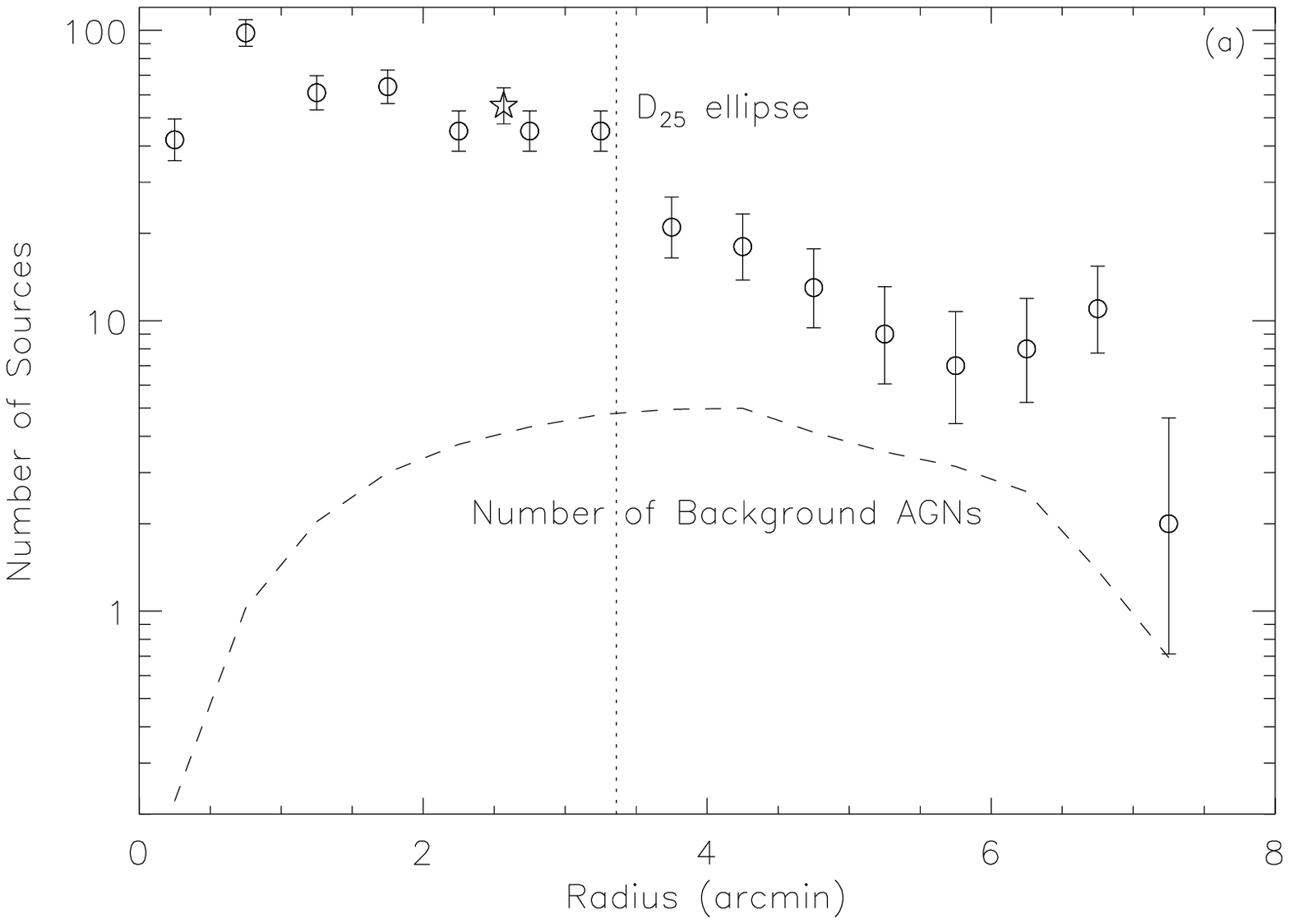}
\includegraphics[scale=0.5]{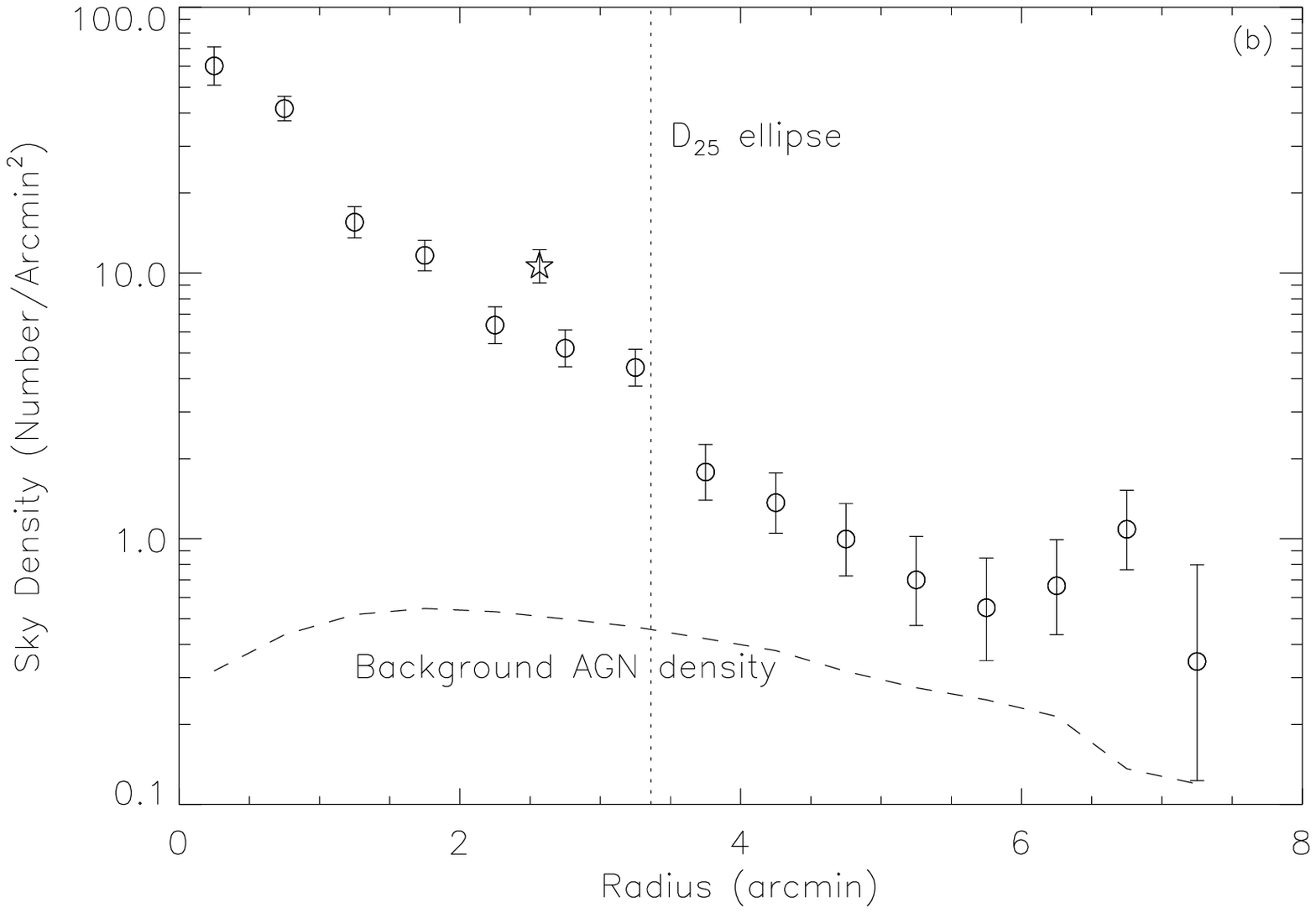}
}
\figcaption{
Radial profile of the (a) number and (b) sky density 
of X-ray sources.
The source numbers and densities were computed in annular regions
centered on the nucleus of the
galaxy; the central 10\arcsec-radius area was excluded.
The star symbol represents the number or density of X-ray sources 
within the D$_{25}$ ellipse of NGC 4647.
The source numbers or densities have not been corrected for 
detection incompleteness
or background AGN contamination.
The dash curve represents the expected numbers or sky densities of 
background AGNs, taking into account the detection
incompleteness. The vertical
dotted line indicates the average radial distance of the NGC 4649 
D$_{25}$ ellipse.
\label{fig-radprofile}}
\end{figure*}

The radial profiles of the number and density of X-ray sources 
are presented in Figure~\ref{fig-radprofile}.
We calculated
X-ray source numbers and densities in annular regions 
centered on the galactic nucleus; the 
central 10\arcsec-radius area was excluded from the calculation.
The 1$\sigma$ uncertainties of the surface densities 
were calculated based on the Poisson
errors of the number of sources in each bin \citep{Gehrels1986}.
We did not correct the source densities for detection incompleteness
or background active galactic nucleus (AGN) contamination. 
We derived the expected background AGN numbers and densities based on the 
\citet{Gilli2007} AGN population-synthesis model, which was 
normalized to 
the AGN surface density observed in the $\approx4$~Ms \chandra\ Deep
Field-South \citep{Xue2011}. The computations of the AGN numbers and densities
also took into account the detection incompleteness, by applying the
positional- and luminosity-dependent
detection probabilities derived in Section~\ref{completeness} above.
There are $\approx45$ background AGNs expected 
among the catalog sources, and $\approx15$ background AGNs within the D$_{25}$ 
ellipse of NGC 4649.
It appears that even at large radius ($\ga7\arcmin$),
there are still some X-ray sources associated with the galaxy, although
the number is limited ($\approx10$).
There are 55 sources within the D$_{25}$ ellipse of the companion galaxy, NGC 4647, which appears to be an overabundance of X-ray sources as indicated in
Figure~\ref{fig-radprofile}b and can be attributed to the 
sources belonging to NGC 4647.
We estimate that $\approx27$ sources belong to NGC 4649, $\approx3$
are background AGNs, and the remaining
$\approx25$ belong to NGC 4647, given the radial profile of the source density.

There are 399 X-ray sources located within the D$_{25}$ ellipse of NGC 4649,
including $\approx15$ background AGNs and $\approx25$ NGC 4647 sources.
The total number of X-ray sources in NGC 4649 is much larger than that
of NGC 3379
or NGC 4278, with 98 sources within
the D$_{25}$ ellipse of NGC 3379 and 180 within the D$_{25}$ ellipse of NGC 4278
\citep{Brassington2008,Brassington2009}.
The \chandra\ observations of NGC 3379 and NGC 4278 are actually deeper
than those of NGC 4649 in terms of the limiting luminosity detected;
the 90\% completeness limit inside the D$_{25}$ ellipse
is $6\times10^{36}$~\lum\ for NGC 3379 \citep{Kim2009},
$1.5\times10^{37}$~\lum\ for NGC 4278 \citep{Kim2009},
and $5.5\times10^{37}$~\lum\ for NGC 4649.
The high density of X-ray sources in NGC 4649
is probably due to the combination of its high optical/IR luminosity and high
GC specific frequency \citep{Boroson2011}; it may also related the
interaction with the companion galaxy, or the past interaction with the
Virgo cluster members.
These differences will be explored more in depth
in D.-W.~Kim et al. (2012, in prep.).

\subsection{Nuclear Source and Ultraluminous X-Ray Sources}

The nuclear X-ray source (XID 253) is 0.24\arcsec\ away from the optical
center of the galaxy.\footnote{This positional offset is dominated by
the astrometric offset between the \chandra\ and optical images; see
Section~\ref{sec-gccp}. There is no significant physical offset
between the X-ray and optical positions.}
It has $\approx870$ FB counts and an observed FB
luminosity of $4.5\times10^{38}$~\lum\ in the merged observation.
The nucleus of NGC 4649 was suggested to host a low-luminosity AGN
powered by a radiatively inefficient
accretion flow \citep[e.g.,][]{Dimatteo1997,Quataert1999}. A central AGN is also
required to produced the observed X-ray cavities \citep{Shurkin2008}.
We fit the AE-extracted spectrum of the nuclear source using XSPEC.
The spectrum cannot be fit with a 
simple absorbed power-law model, and there
is clearly a strong soft X-ray excess around 1~keV,
which is typical among low-luminosity AGNs
and is considered to originate from hot gas in the galactic nucleus
\citep[e.g.,][]{Ptak1999}.
We thus fit the spectrum with an
absorbed power-law (AGN) plus thermal plasma (hot gas) model
({\sc wabs1*pow+wabs2*apec}). The absorption
column density for the thermal component was fixed at the Galactic value,
and the plasma temperature, power-law photon index, and intrinsic absorption
are free parameters. The resulting best fit is statistically 
acceptable ($\chi^2/$dof=0.93 
and null hypothesis probability $\approx0.6$),
with temperature $T=1.3_{-0.1}^{+0.2}$~keV, photon index $\Gamma=1.9_{-0.6}^{+0.5}$,
and intrinsic absorption $N_{\rm H,int}=0.2_{-0.2}^{+1.3}\times10^{22}$~cm$^{-2}$.
The errors are at the 90\% confidence level for one parameter of interest.
The intrinsic 2.0--8.0 keV X-ray luminosity for the power-law component is
$1.5\times10^{38}$~\lum\ after absorption correction, indicating
its low-luminosity nature.
This source is also variable based on the chi-square or flux variation test
below (see Section \ref{sec-var}). The power-law photon index of $\approx2$,
moderate intrinsic absorption, and long-term variability confirm the nature
of the nuclear source as a low-luminosity AGN \citep[e.g.,][]{Turner1997,Risaliti2002}.

There are nine sources (XIDs 73, 81, 106, 152, 171, 392, 421,
422, and 501) with ULX luminosities ($L_{\rm 0.3\textrm{--}8~keV}>10^{39}$~\lum)
in the merged observation
or one of the individual observations. 
It is probable that some of these sources are 
background AGNs instead of ULXs, as we expect
$\approx1.7$ background AGNs
with ULX fluxes if placed at the distance of
NGC 4649.
In fact, XIDs 73 and 501 are located $\approx7\arcmin$
away from the nucleus, and thus have a 
higher chance of being background AGNs.
Moreover, XID 152 was identified as a foreground star based on
optical observations (see Section~\ref{sec-gccp} below).

\subsection{Hardness Ratios and X-Ray Colors}

We calculated the HRs and X-ray colors of 
the sources to characterize their spectral properties.
The X-ray HR is defined as HR=$(C_{\rm HB}-C_{\rm SB})/(C_{\rm HB}+C_{\rm SB})$,
and the X-ray colors are defined as
SC=$(C_{\rm SB2}-C_{\rm SB1})/C_{\rm FB}$ (soft color) and 
HC=$(C_{\rm HB}-C_{\rm SB2})/C_{\rm FB}$ (hard color), where
$C_{\rm FB}$, $C_{\rm SB}$, $C_{\rm HB}$, $C_{\rm SB1}$, and $C_{\rm SB2}$
are the source count rates in the FB, SB, HB, SB1, and SB2, 
 which have been corrected for Galactic 
absorption. The definition of the X-ray bands, HR, and colors are
summarized in Table~\ref{tbl-bands}.
To better constrain the HRs and colors and their associated 
errors in the low-count regime, we adopted the Bayesian approach 
developed by \citet{Park2006}.
This approach provides a rigorous statistical treatment of the Poisson
nature of the detected photons as well as the non-Gaussian nature of
the error propagation, and it takes directly the AE extracted source counts,
background counts, and appropriate scaling factors as input parameters.

\begin{figure}
\centerline{
\includegraphics[scale=0.5]{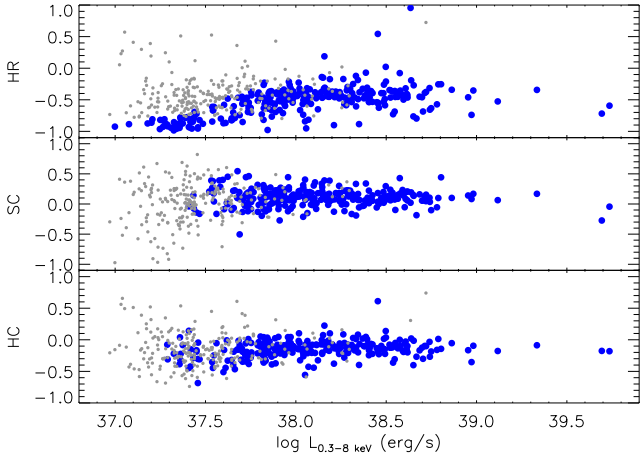}
}
\figcaption{
Luminosity vs. HR (top panel), 
luminosity vs. soft color (middle panel), 
and luminosity vs. hard color (bottom panel) for all the sources detected
in the merged observation.
Blue data points represent sources with relatively small HR or color errors 
(smaller than
the $3\sigma$-clipped mean of the errors for all the sources),
while
gray data points represent sources with large errors and are less significant.
In the top panel, there are some blue data points in the low-luminosity 
regime which do not appear in the other two panels. These are
extremely soft sources with $\approx10$--30 detected counts in the SB
and almost zero count in the HB, and thus their HRs were constrained 
to be very close to $-1$ with relatively small errors while their X-ray colors
still have large uncertainties.
(A color version of this figure is available in the online journal.)
\label{propco}}
\end{figure}

\begin{figure}
\centerline{
\includegraphics[scale=0.5]{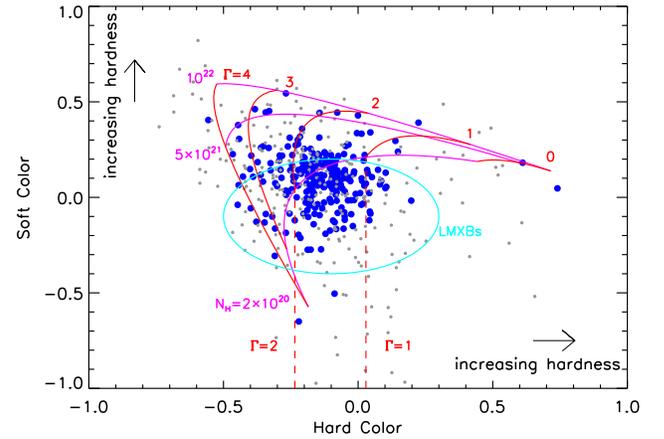}
}
\figcaption{
X-ray color--color plot for the catalog sources.
Blue data points represent sources with relatively small color errors 
(smaller than
the $3\sigma$-clipped mean of the errors for all the sources), 
while
gray data points represent sources with large errors and are less significant.
The solid red and magenta tracks show the expected X-ray colors 
from absorbed power-law
spectra with different power-law indices and column densities.
The two dashed red lines represent the 
expected hard colors from
unabsorbed power-law models with $\Gamma=1$ and $\Gamma=2$; data points in
between of these lines are likely soft-excess sources.
The cyan ellipse indicates the area that is likely dominated by LMXBs
\citep{Prestwich2003}.
A significant fraction
of the X-ray sources with small color errors are located in
the region expected to be dominated by LMXBs.
(A color version of this figure is available in the online journal.)
\label{colorplot}}
\end{figure}

The luminosity--HR, luminosity--soft color, and luminosity--hard color 
plots are presented
in Figure~\ref{propco}. 
There is no significant dependence of the HR or X-ray color
on the X-ray luminosity.
The X-ray 
color--color plot is shown in Figure~\ref{colorplot}.
The red and magenta tracks (solid curves) show
the expected X-ray colors from absorbed power-law
spectra with different power-law indices and column densities.
Sources outside the area enclosed by these tracks,
are mostly soft-excess sources, of which the soft and hard colors
cannot be simultaneously explained by an absorbed power-law spectrum.
The soft excess probably originates from thermal gas emission if
the source is close to the nucleus. For these soft-excess sources,
we show 
two vertical red lines indicating the expected hard colors from
unabsorbed power-law models with $\Gamma=1$ and $\Gamma=2$ (if absorption
is present, the power-law index will be larger for a given hard color).
The X-ray color--color plot can be used to separate
the X-ray sources into groups that are likely
dominated by certain source types 
\citep[e.g.,][]{Colbert2004,Prestwich2003,Prestwich2009}.
In Figure~\ref{colorplot}, 
the cyan ellipse indicates the area that is likely dominated by LMXBs
\citep{Prestwich2003}.
A significant fraction ($\approx75\%$)
of the X-ray sources are located in the region dominated by LMXBs, 
most of which have $\Gamma=1.5$--2.0 and no/little intrinsic absorption, 
as expected for 
the X-ray source population in an early-type galaxy \citep[e.g.,][]{Fabbiano2006}.

\subsection{Source Variability and Transient Candidates} \label{sec-var}

X-ray flux/spectral variability is a common feature among XRBs in
galaxies, which is generally attributed to the change of physical
properties of the accretion disks
\citep[e.g.,][]{Done2007,Brassington2010,Fabbiano2010}.
The six \chandra\ observations of NGC 4649 span $\approx11$ years,
allowing us to study the long-term variability of the X-ray sources
and search for transient candidates.

\begin{figure}
\centerline{
\includegraphics[scale=0.5]{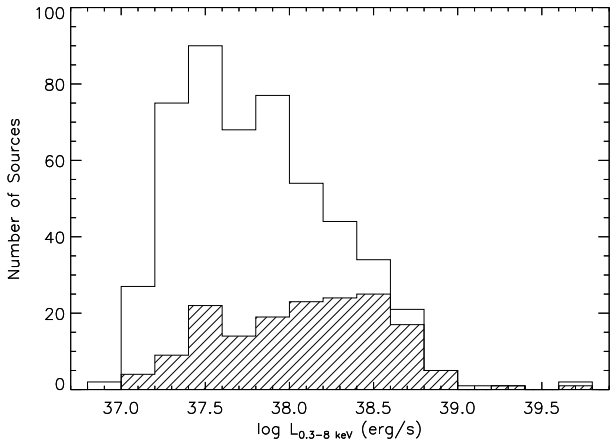}
}
\figcaption{FB luminosity distributions
of all the 501 sources (unshaded histogram) and the 164 variable sources
(shaded histogram) in the merged observation; 
3$\sigma$ upper limits on the luminosities were
used for the 8 undetected sources.
Note that for
low-luminosity sources which generally have large uncertainties in
the observed luminosities, the data
are probably not able to reveal their variabilities.
\label{lhistplot}}
\end{figure}

We define long-term source variability using the chi-square test
described in \citet{Brassington2009}. For every source, we performed 
least squares fitting
to the FB luminosities observed in the six individual observations 
with a flat line model. 
In cases where the source is not detected (but still covered by the 
observation), we set the luminosity to be the 1$\sigma$ upper 
limit with the same value as the errors.
If the reduced chi-square value of the best-fit model is greater than 1.2
($\chi_{\rm red}^2>1.2$), the source is determined to be variable;
otherwise, it is non-variable. Of the 501 sources, 164 are variable, 
331 are non-variable.
The other six sources are covered by only
one observation, and their long-term variabilities were not constrained. 
In Figure~\ref{lhistplot}, we show the FB luminosity distributions
of all the 501 sources (unshaded histogram) and the 164 variable sources
(shaded histogram).
We note that for sources covered by only a few observations or 
low-luminosity sources that have large uncertainties in
the observed luminosities, the data
are probably not able to reveal their variabilities.
The fraction of variable sources in NGC 4649 (33\%) is slightly smaller than
that in NGC 3379 (42\%) or NGC 4278 (44\%). The difference may
be partially caused by the different methods adopted to 
define source detections and calculate errors and upper limits, and it
may be also related to the different depths probed by \chandra\ in these
galaxies.

Besides the chi-square test, we further investigated the variation 
of the source fluxes by comparing the FB count rates between observations.
For a source that was detected in at least one individual observation and
was covered by at least two observations, we
computed the maximum statistical significance of its flux variation between
any two observations, defined as
\citep[e.g.,][]{Brassington2009,Sell2011}:
\begin{equation}
\sigma_{\rm var}={\rm max}_{i,j}\frac{|{C}_i-{C}_j|}{\sqrt{\sigma_{{C}_i}^2+\sigma_{{C}_j}^2}}~,
\end{equation}
where the subscripts $i$ and $j$ run over different observations, and ${C}_i$,
${C}_j$, $\sigma_{{C}_i}$, and $\sigma_{{C}_j}$ are the count rates 
(or 3$\sigma$ upper limits if not detected)
and their associated 1$\sigma$ errors.
We consider a source to be variable if its $\sigma_{\rm var}$
parameter is greater than three (i.e., $>3\sigma$ variation).
The variation significances are list in Table~\ref{tbl-main}; 49 sources
are determined to be variable, all of which are also variable based on the
chi-square test.

\begin{figure*}
\centerline{
\includegraphics[scale=0.5]{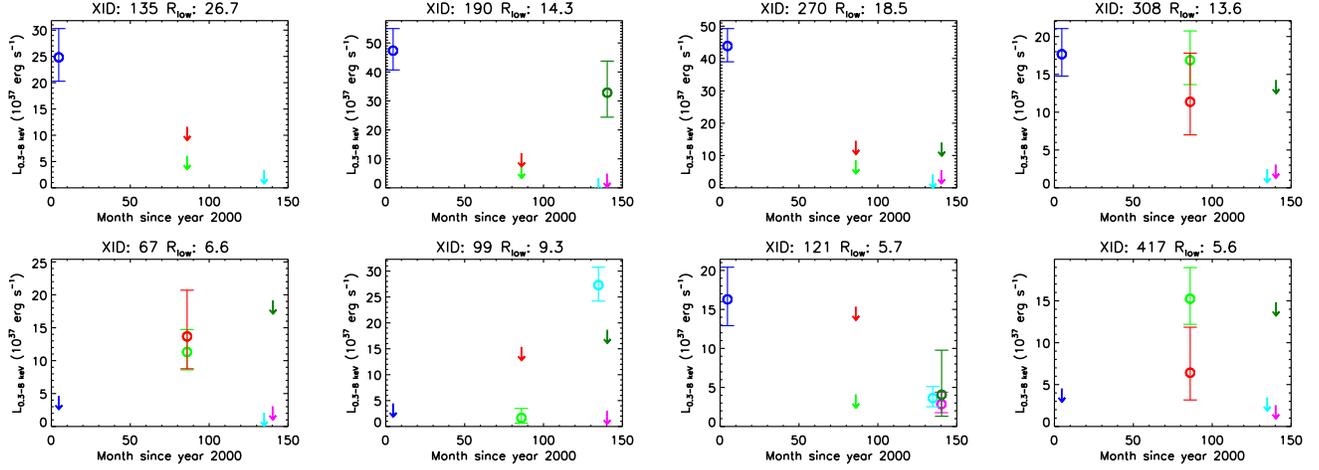}
}
\figcaption{
Light curves of the four transient candidates
(XIDs 135, 190, 270, and 308)
and four potential transient candidates (XIDs 67, 99, 121, and 417).
The associated 1$\sigma$ errors or 3$\sigma$ upper limits (if undetected) for the 
FB luminosities are shown
(see Section 3.1 for details).
The lowest values of the lower bounds of the ratios between the
``on-state'' and ``off-state'' count rates are indicated.
Date points are color coded for different observations: the blue,
green, red, cyan, magenta, and 
dark green colors represent observations 1--6 respectively.
(A color version of this figure is available in the online journal.)
\label{tcplot}}
\end{figure*}

We searched for transient candidates following \citet{Brassington2009}.
For a source detected in one observation but not another, we 
calculated its 1$\sigma$ lower bound of the ratio between the 
``on-state'' and ``off-state'' count rates, utilizing the \citet{Park2006}
Bayesian approach and the AE extraction results.
Such lower bounds of the ratios were calculated for all available pairs of 
observations. 
The source is considered as a transient candidate
if the lowest value among all the lower bounds
is greater than 10, or a potential transient 
candidate if the lowest value is between 5 and 10. We discovered
four transient candidates (XIDs 135, 190, 270, and 308) and 
four potential transients
(XIDs 67, 99, 121, and 417), and their light curves
are displayed in Figure~\ref{tcplot}. As shown in the plot, 
four of these objects 
have a maximum luminosity $>2\times10^{38}$~\lum, likely being
black-hole XRBs, and the other four all have a maximum luminosity 
$>10^{38}$~\lum.
All these eight sources are labeled
as variable based on the chi-square or flux variation test.
Note that strongly variable sources are not identified as transients if 
they are detected in all the observations. There are eight sources 
with $\sigma_{\rm var}>7$ (XIDs 70, 99, 152, 171, 235, 270, 421, and 486) in 
the catalog,
six of which are not transient candidates; these six heavily variable sources
are very luminous ($>2\times10^{38}$~\lum).

Short-term variability for each source was examined when it has more than
20 FB counts in a single observation. We ran the Kolmogorov-Smirnov test
to search for variability in the FB count rate and 
visually inspected the light curves of every
candidate. After excluding sources showing only variability at the beginning
or the end of an observation, we found one source, XID 421, that has
a significant short-term variation in count rate. It is a ULX 
($L_{\rm 0.3\textrm{--}8~keV}=2.2\times10^{39}$~\lum) and also exhibits a significant long-term
variability ($\sigma_{\rm var}$=14.4).
This source is discussed in more details in \citet{Roberts2012}.

\begin{figure}
\centerline{
\includegraphics[scale=0.45]{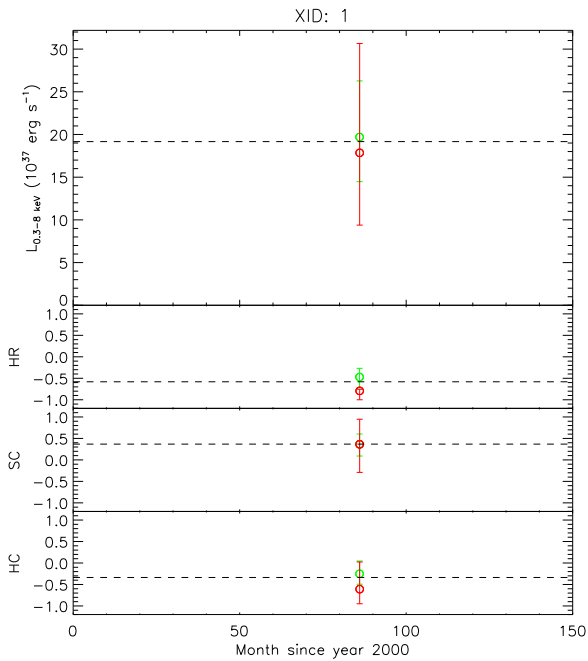}
}
\figcaption{
The FB luminosity (top panel), HR (second panel), soft color (third panel), 
and hard color (bottom panel) of each source
as a function of the observation date. The associated 1$\sigma$ errors 
are plotted; for undetected sources, the 3$\sigma$ upper limits on the
luminosities are shown. 
Date points are color coded for different observations: the blue,
green, red, cyan, magenta, and 
dark green colors represent observations 1--6 respectively.
In each panel, the dashed line indicates the 
value derived from the merged observation. As the count-rate-to-flux conversion factors
were derived individually (Section 3.1), 
the luminosities for the merged and individual observation
may differ slightly even if the source is covered by only one observation.
(An extended and color version of this figure is available in the online journal.)
\label{propindplot}}
\end{figure}

\begin{figure}
\centerline{
\includegraphics[scale=0.45]{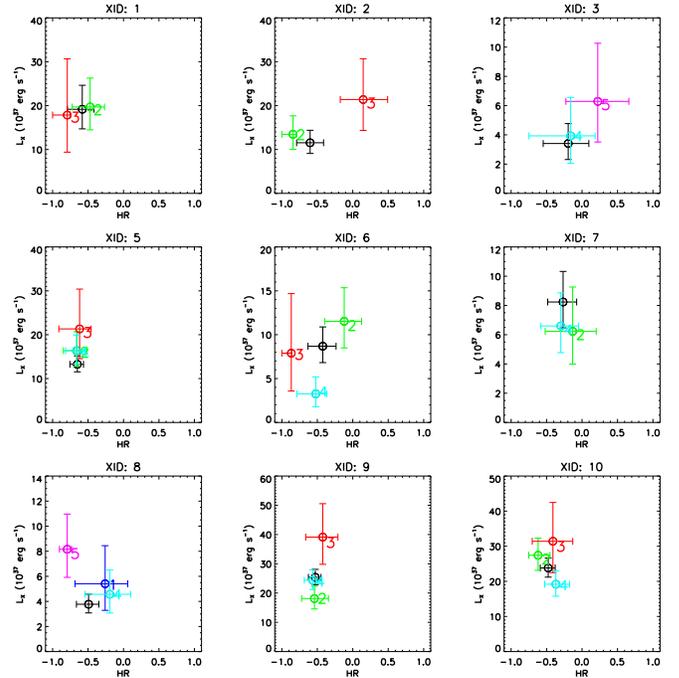}
}
\figcaption{
FB luminosity vs. HR for sources in the merged and individual observations.
Only sources detected in the merged observation are displayed.
The 1$\sigma$ errors are shown for the luminosities and HRs.
Date points are color coded for different observations: the blue,
green, red, cyan, magenta, and
dark green colors represent observations 1--6 respectively, and 
the black symbol indicates the merged observation.
For source covered by only one
observation, the merged data point overlaps with the one for the individual observation.
(An extended and color version of this figure is available in the online journal.) 
\label{hrlindplot}}
\end{figure}

\begin{figure}
\centerline{
\includegraphics[scale=0.45]{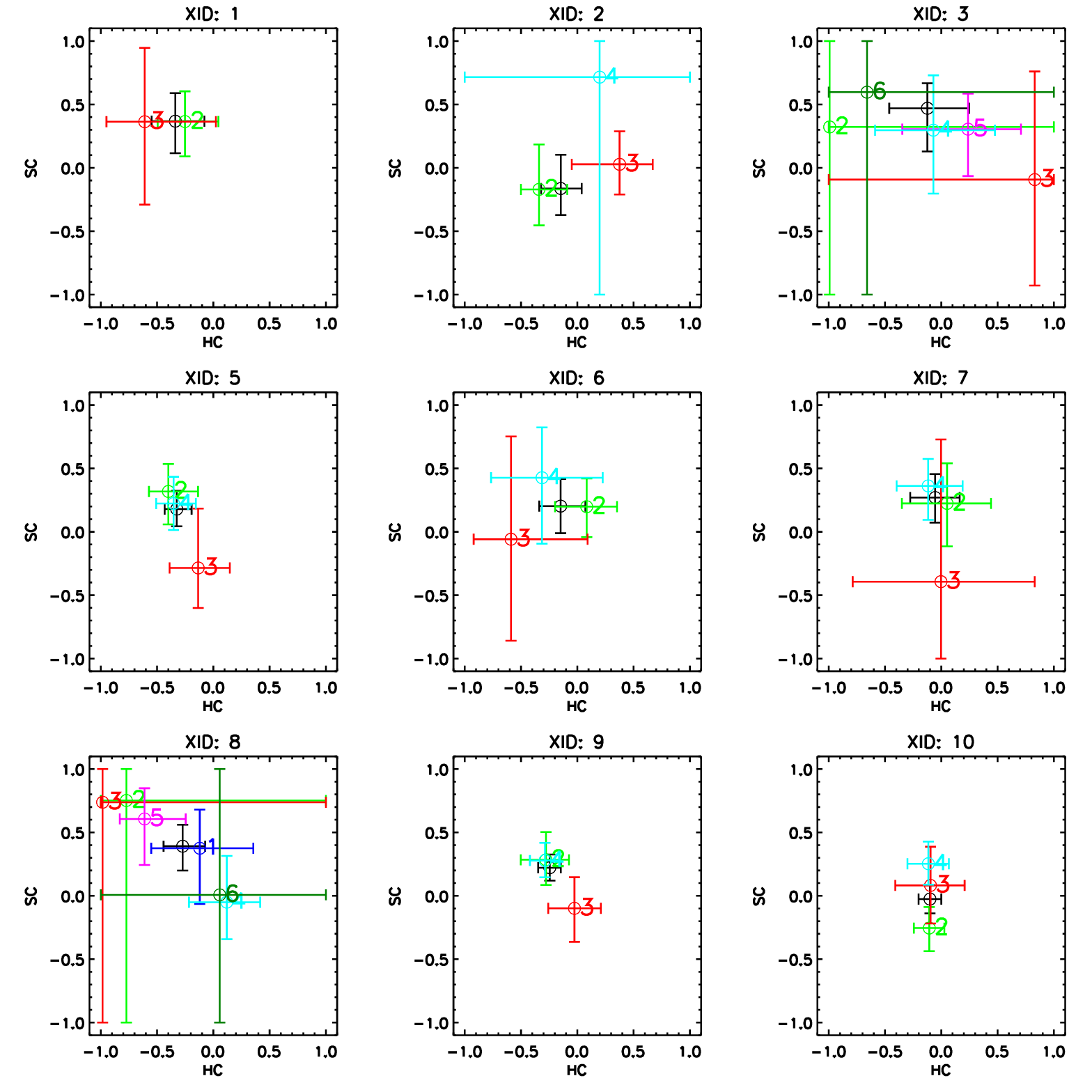}
}
\figcaption{
Soft color vs. hard color for sources in the 
merged and individual observations.
Only sources detected in the merged observation are displayed.
The 1$\sigma$ errors are shown for the colors.
Date points are color coded for different observations: the blue,
green, red, cyan, magenta, and
dark green colors represent observations 1--6 respectively, and
the black symbol indicates the merged observation. For source covered by only one
observation, the merged data point overlaps with the one for the individual observation.
(An extended and color version of this figure is available in the online journal.)
\label{colorindplot}}
\end{figure}

Furthermore, we explored source spectral variabilities.
In Figure~\ref{propindplot}, we present the FB luminosities, 
HRs, and X-ray colors of each source in all six observation.
In Figures~\ref{hrlindplot} and \ref{colorindplot}, 
we show the luminosity--HR and color--color
plots for each source in individual observations.
It is evident that the spectral properties of the sources 
vary frequently beyond the 1$\sigma$ bounds, 
and both high/soft--low/hard and low/soft--high/hard 
spectral transitions are present,
as have been observed in 
NGC 3379 and NGC 4278 \citep{Brassington2008,Brassington2009}.

\subsection{Optical Counterparts} \label{sec-gccp}

We matched the X-ray sources to the GC catalog of NGC 4649 obtained
via {\it Hubble Space Telescope}/Advanced Camera for Surveys ({\it HST}/ACS) 
observations
\citep{Strader2012}. We first checked the 
astrometric transformation between the X-ray and optical images based 
on an initial list of high-probability matches between X-ray sources 
and GCs. We found a small but significant 
systematic offset ($0.28\arcsec$ in right ascension and $0.04\arcsec$
in declination)
between the \chandra\ and {\it HST} astrometry, which we corrected before matching. There was no evidence for significant higher-order terms in the 
astrometric transformation. 
Note that the source positions presented in this paper
follow the \chandra\ astrometry.

Most putative matches were within 0.5\arcsec; we thus 
used 0.6\arcsec\ as our limit 
for reliable X-ray to GC matches. 
There are 157 matches by this criterion. 
A further 4 objects have offsets between $0.6\arcsec\textrm{--}0.8\arcsec$; we 
consider these as possible matches.

There are X-ray sources outside of the field of view of the {\it HST}/ACS
mosaic presented in \citet{Strader2012}. 
We thus searched for matches between 
these sources and the ground-based photometric GC catalog of 
\citet{Lee2008}, using the same astrometric criteria as for the {\it HST} 
matches. We found 12 X-ray sources that appear to be associated with 
these photometric GCs. All matches are with relatively luminous clusters. These are considered as reliable matches, but because these GC candidates are 
not resolved, it is possible that a few of them are background contaminants.

Of the X-ray sources within the {\it HST}/ACS mosaic that do not match to
GCs, many are nonetheless associated with an optical source. 17 are candidate 
background galaxies for which the sources are likely to be AGNs. 
One other source (XID 144) matches an unusual dwarf galaxy that is 
discussed in a separate paper (J.~Strader et al. 2012, in prep.).
In two cases the matches are with relatively luminous point sources that appear to be foreground stars. One optical counterpart is the nucleus of 
NGC 4649. 
As discussed in Section~\ref{radprofile}, we expect $\approx25$ sources
belonging to the companion galaxy NGC 4647.
Based on the {\it HST}/ACS and Sloan Digital Sky Survey (SDSS; \citealt{York2000}) images, we selected 35 X-ray sources that are located where
the underlying optical light is dominated by NGC 4647
(ratio of NGC 4647 to NGC 4649 light is very high). We further identified
8 sources located where the optical light is likely 
dominated by NGC 4647 (these 8 sources are around the edge of the 
D$_{25}$ ellipse of NGC 4647). These 35 plus 8 sources are the likely
candidates for the expected $\approx25$ NGC 4647 sources; they also have
the chance of being associated with NGC 4649 or being a background AGN.
The remaining 264 X-ray sources 
($53\%$ of the total) are not associated with any 
obvious optical source, 201 of which are within the 
D$_{25}$ ellipse and most of these sources
are likely to be LMXBs in the field of 
NGC 4649. 

There are 173 sources in total that have a GC counterpart, 
and their positions are shown in Figure~\ref{fig-pos};
141 (82\%) of these objects are within the D$_{25}$ ellipse.
Considering those X-ray sources within the D$_{25}$ ellipse
and outside 10\arcsec\ of the galactic center, the fraction of 
GC-LMXBs is $36\%$ (140/387). This value is in between of 
the GC-LMXB fractions found 
in NGC 3379 (24\%) and NGC 4278 (47\%; \citealt{Kim2009}),
likely 
consistent with previous findings \citep[e.g.,][]{Juett2005,Kim2006,Kim2009} 
that 
the GC-LMXB fraction increases with increasing GC specific frequency
(1.2, 6.9, and 5.2 for NGC 3379, NGC 4278, and NGC 4649, respectively;
\citealt{Boroson2011}). We caution that the completeness limit differs
among these galaxies (see Section~\ref{radprofile}), and thus the 
GC-LMXB fractions may not be directly comparable; the relation between the
GC-LMXB fraction and GC specific frequency will be explored in more detail in
D.-W.~Kim et al. (2012, in prep.).

\begin{figure*}
\centerline{
\includegraphics[scale=0.5]{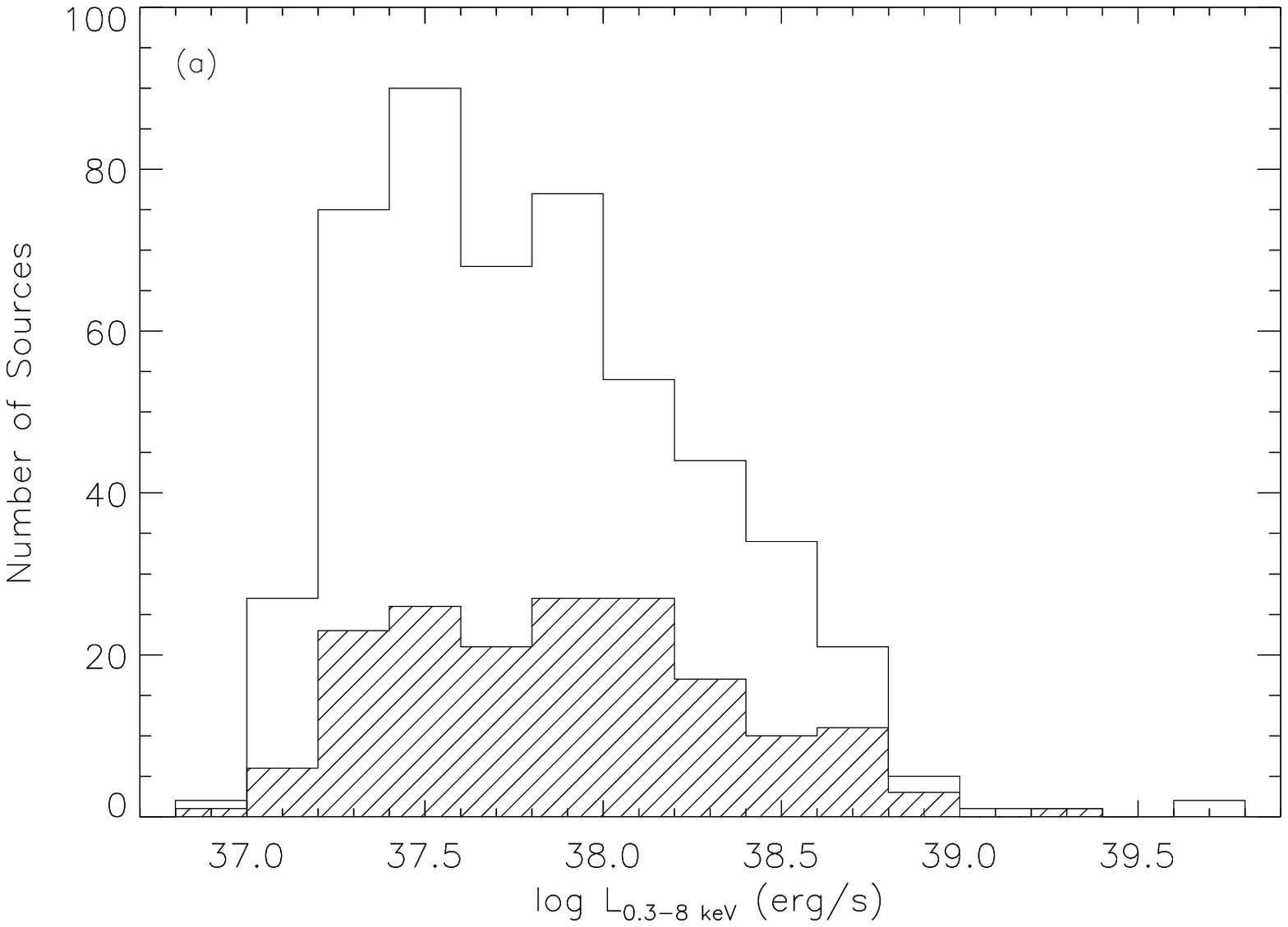}
\includegraphics[scale=0.5]{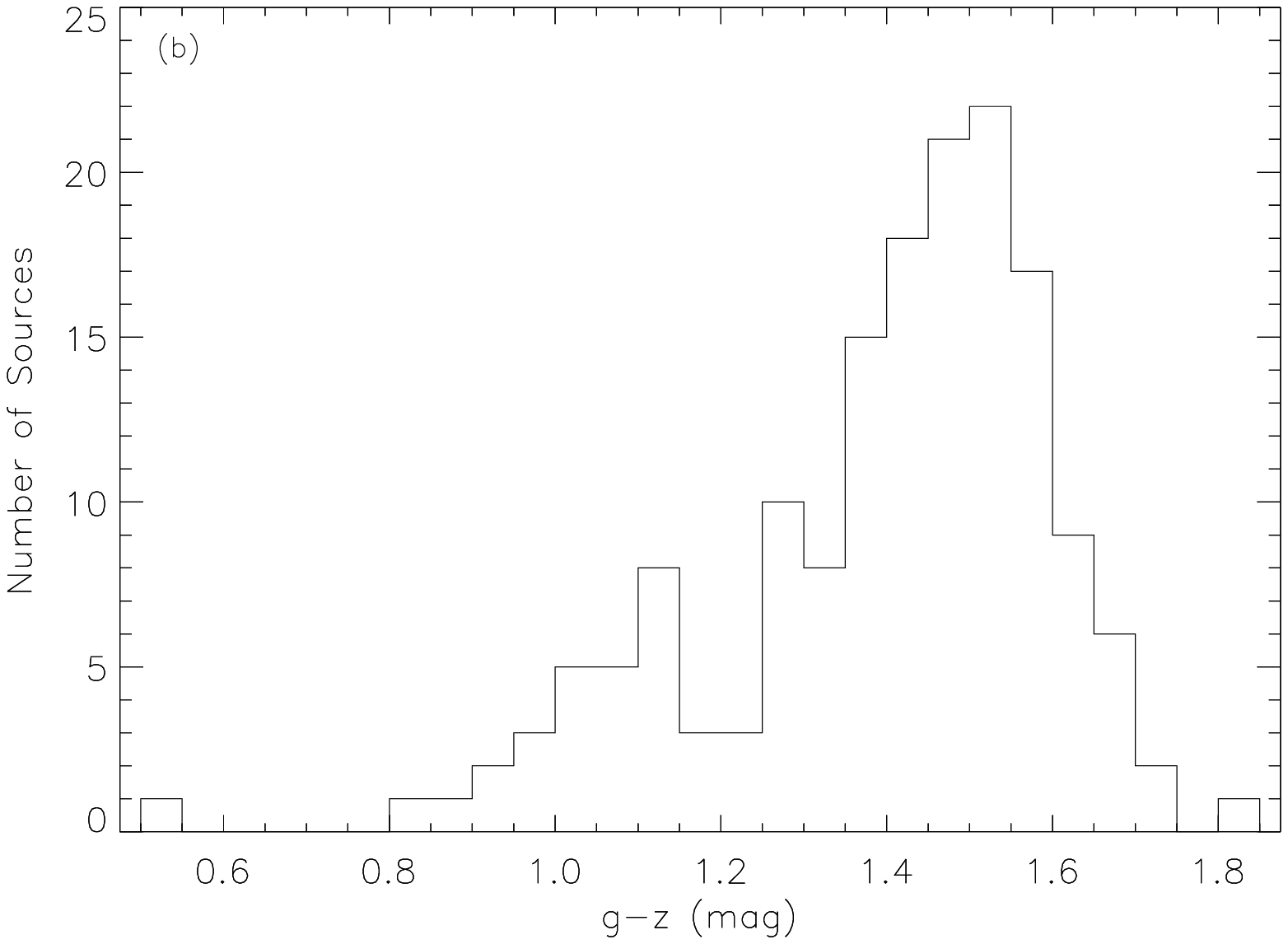}}
\caption{(a) FB luminosity distributions
of all the 501 sources (unshaded histogram) and the 173
GC-LMXB sources
(shaded histogram) in the merged observation;
3$\sigma$ upper limits on the luminosities were
used for the 8 undetected sources. (b) {\it HST} $g-z$ color distribution
for the 161 GC-LMXB associations from \citet{Strader2012}.
LMXBs tend to be hosted by red GCs.
\label{gclhistplot}}
\end{figure*}

The GC-LMXB associations are slightly more X-ray luminous than the 
entire X-ray sample on average,
as shown in Figure~\ref{gclhistplot}a (median luminosity
$7.4\times10^{37}$~\lum\ vs. $5.7\times10^{37}$~\lum). 
The relative lack of low-luminosity GC-LMXBs 
when compared with field LMXBs
was previously reported
and discussed in \citet{Kim2009}, which may be an 
intrinsic feature of the LMXB populations.
The fraction of variable sources among GC-LMXBs (34\%)
is comparable to that for the entire sample (33\%).

None of the eight transient or potential
transient candidates is matched to a GC.
For the nine sources with ULX luminosities, one (XID 152) 
was identified as a foreground star,
and another four (XIDs 81, 171, 392, and 421) have a secure GC counterpart.
We investigated the color distribution of the GC-LMXBs, using
the $g-z$ colors ({\it HST} F475W and F850LP filters)
for the 161 GC-LMXB associations obtained from \citet{Strader2012}.
The color histogram is displayed in Figure~\ref{gclhistplot}b.
The median color value is 1.44
with an interquartile range of 1.29--1.54.
Therefore, the LMXBs in NGC 4649 
are preferentially hosted by red GCs, consistent with previous
findings of the GC-LMXB connection in other galaxies, and
likely indicating the importance of metallicity in the
formation of GC-LMXBs 
\citep[e.g.,][]{Sarazin2003,Jordan2004,Kim2006,Paolillo2011}.

\subsection{Source Catalog} \label{sec-catalog}

Photometric properties for the 501 X-ray sources are presented in
Tables~\ref{tbl-main}--\ref{tbl-ind6}.
The details of the Table~\ref{tbl-main} columns are listed below.

\begin{enumerate}

\item
Column~1: the X-ray source identification number (XID). 
Sources are listed in order of increasing right ascension.
\item
Column~2: the source name following the IAU convention (CXOU Jhhmmss.s$+/–$ddmmss).
\item
Columns~3 and 4: the right ascension and declination of
the \hbox{X-ray} source, respectively.

\item
Column~5: the radial distance of
the source to the nucleus of NGC 4649, in units of arcminutes.

\item
Column~6: the source positional uncertainty at the 95\% confidence level, 
in units of arcseconds (see Section 3.1).

\item
Column~7: the logarithm of the observed FB luminosity, 
in units of erg s$^{-1}$. 
A 3$\sigma$ upper limit is given if the source is not detected in the FB.

\item
Column~8: the long-term variability flag (see Section 3.5). 
The source is labeled as variable (``V''), non-variable (``N''),
transient candidate (``TC''), or potential transient candidate (``PTC'').
All transient candidates are variable.

\item
Column~9: the maximum statistical significance of the FB 
flux variation between any two observations (see Section 3.5).
It is set to ``$-1.0$'' if the source is 
not covered by at least two observations or not detected in at least one observation. 

\item
Column~10: the positional flag. The source may be located within
the D$_{25}$ ellipse of NGC 4649 (``1''), 
within the D$_{25}$ ellipse of NGC 4647 (``2''),
within both D$_{25}$ ellipses (``3''), or outside the ellipses (``0'').

\item
Column~11: the note on the optical counterpart. The source may have 
a reliable {\it HST} GC counterpart (``1''; 157 sources),
a probable {\it HST} GC counterpart (``2''; 4 sources),
a ground-based GC counterpart (``3''; 12 sources), 
a background AGN counterpart (``4''; 17 sources),     
a counterpart that is an unusual dwarf galaxy (``5''; 1 source),
a counterpart that is the nucleus of NGC 4649 (``6''; 1 source),
a counterpart that is a foreground star (``7''; 2 sources),
or it has a high chance ($>50\%$) of being associated with the companion galaxy
NGC 4647 (``8''; 35 sources), it has a 
less significant chance of being associated
with NGC 4647 (``9''; 8 sources), or it is associated with NGC 4649
but having no counterpart (``0''; 264 sources).

\item
Column~12: the GC ID from \citet{Strader2012} or Lee et al. (2008; starting with the letter ``L'') for the 173 X-rays sources with a 
GC counterpart.

\end{enumerate}

In Table~\ref{tbl-mainadd}, we list source properties including 
the net counts in the five bands (Section 3.1), 
HRs (Section 3.4), and X-ray colors (Section 3.4) and their associated 1$\sigma$
errors in 
the merged observation. In Tables~\ref{tbl-ind1}--\ref{tbl-ind6}, we
present these properties in the six individual observations. Note that 
a single observation does not cover all the sources.

\section{Summary}
We have presented a catalog and basic analyses of the X-ray sources detected 
in NGC 4649. The key results are summarized in the
following:

\begin{enumerate}
\item
NGC 4649 has been covered by six \chandra\ ACIS-S3 observations, with a total
cleaned exposure of 299.4~ks.

\item
The \chandra\ source catalog consists of 501 sources that were
detected following a two-step source-detection
approach using {\sc wavdetect} and AE. First, 517 candidate sources were detected
in the merged and individual FB images using {\sc wavdetect} with 
a false-positive probability threshold
of $1\times10^{-6}$. Then we filtered out 16 less-significant
candidates using the AE no-source probability parameter. The resulting
source catalog is highly reliable.

\item
The source photometry was extracted by AE, using polygonal 
source-count extraction regions to approximate the shape of the PSF.
In order to study source variability, photometry was derived for 
the merged observation as well as the six individual
observations.
In the merged observation, the number of FB net counts ranges from
$\approx8$ to $\approx3870$, and the FB luminosity ranges from
$9.3\times10^{36}$~\lum\ to $5.4\times10^{39}$~\lum.

\item
We performed simulations to assess the source detection completeness.
The average 50\% (90\%) completeness limit of the D$_{25}$ region
is $2.1\times10^{37}$ ($5.5\times10^{37}$)~\lum.
There are $\approx45$ ($\approx15$)
background AGNs expected
among the catalog sources (within the D$_{25}$ ellipse).
Of the 501 catalog sources, 399 are located within the D$_{25}$ ellipse of NGC 4649,
and 55 within the D$_{25}$ ellipse of NGC 4647. 
We estimate that $\approx25$ sources are associated with
NGC 4647, given the radial profile of the source density.
NGC 4649 hosts a larger population of X-ray sources compared to NGC 3379 or NGC 4278.

\item The nuclear source is a low-luminosity AGN, with an intrinsic 2.0--8.0
keV X-ray luminosity of $1.5\times10^{38}$~\lum. The spectrum also shows a 
thermal component from the nuclear hot-gas emission at $\approx1.3$~keV. There are 
nine sources with ULX luminosities.

\item
We derived HRs and X-ray colors for the catalog sources in 
the merged observation and the six individual
observations, adopting a Bayesian approach to deal with
the Poisson
nature of the detected photons as well as the non-Gaussian nature of
the error propagation. In the X-ray color--color plot, 
a significant fraction ($\approx75\%$)
of the X-ray sources are located in the region dominated by LMXBs,
with $\Gamma=1.5$--2.0 and no/little intrinsic absorption.

\item
We investigated the long-term variability of the X-ray sources. The chi-square
test revealed 164 variable sources, 49 of which have more than $3\sigma$ variation
in observed fluxes. We identified four transient candidates and four potential
transient candidates based on the ratio of the count rates. X-ray spectral variabilities
are also present among the catalog sources.

\item
We identified 173 GC-LMXB associations based on {\it HST}
and ground-based data. 
These GC-LMXBs appear to be slightly more X-ray luminous than the
entire X-ray sample on average, and the fraction of variable GC-LMXBs
is comparable to that for the entire sample.
The GC-LMXBs tend to have red $g-z$ colors.

\end{enumerate}

~\\

We acknowledge financial support from NASA
{\it HST} grant GO-12369.01-A (BL, GF),
CXC grant GO1-12110X (BL, GF),
and NSF grant AST-0808099 (JPB).
We acknowledge support from the CXC, which is operated by the
Smithsonian Astrophysical Observatory (SAO) for and on behalf of
NASA under Contract NAS8-03060.
GF and TF thank the Aspen Center for Physics.
TF acknowledges support from the CfA and the ITC prize fellowship programs,
and JSG thanks the University of Wisconsin Graduate School and College 
of Letters Science for their research support.
We thank the referee for carefully
reviewing the manuscript and providing constructive comments.


\begin{deluxetable}{cccccl}
\tablewidth{0pt}

\tablecaption{\chandra\ Observations of NGC 4649}
\tablehead{
\colhead{Obs. \#}                                 &
\colhead{Obs. ID}                                 &
\colhead{Start Date}                                 &
\colhead{Exp (ks)}                             &
\colhead{Cleaned Exp (ks)}                         &
\colhead{PI} \\
\colhead{(1)}         &
\colhead{(2)}         &
\colhead{(3)}         &
\colhead{(4)}         &
\colhead{(5)}         &
\colhead{(6)}
}
\startdata
1&785 &2000 Apr 20 &37.4 & 34.2 & C. L. Sarazin\\
2&8182 &2007 Jan 30  &53.0 & 49.2 &P. Humphrey\\
3&8507 & 2007 Feb 1 &  17.8 & 17.3&P. Humphrey\\
4&12976 &2011 Feb 24 & 102.4 & 100.3 & G. Fabbiano\\
5&12975 &2011 Aug 8 & 86.1 & 84.4 & G. Fabbiano\\
6&14328 &2011 Aug 12 & 14.2 & 14.0 & G. Fabbiano\\
\enddata
\tablecomments{
Column 1: observation number, in order of the start date.
Column 2: \chandra\ observation identification number.
Column 3: observation start date.
Column 4: nominal exposure time.
Column 5: exposure time after removing background flares.
Column 6: name of the Principal Investigator.}
\label{tbl-obs}
\end{deluxetable}

\begin{deluxetable}{ll}
\tablewidth{0pt}
\tablecaption{Definition of Energy Bands and X-Ray Colors}
\tablehead{
\colhead{Band} &
\colhead{Definition}   \\
}
\startdata
Full band (FB) &0.3--8.0 keV \\
Soft band (SB) & 0.3--2.0 keV \\
Hard band (HB) & 2.0--8.0 keV \\
Soft band 1 (SB1) & 0.3--1.0 keV \\
Soft band 2 (SB2)& 1.0--2.0 keV \\
Hardness ratio (HR) &   $(C_{\rm HB}-C_{\rm SB})/(C_{\rm HB}+C_{\rm SB})$ \\
Soft X-ray color (SC) & $(C_{\rm SB2}-C_{\rm SB1})/C_{\rm FB}$ \\
Hard X-ray color (HC) & $(C_{\rm HB}-C_{\rm SB2})/C_{\rm FB}$ \\
\enddata
\tablecomments{
$C_{\rm FB}$, $C_{\rm SB}$, $C_{\rm HB}$, $C_{\rm SB1}$, and $C_{\rm SB2}$
are the source count rates in the FB, SB, HB, SB1, and SB2.}
\label{tbl-bands}
\end{deluxetable}

%
%

\begin{deluxetable}{lccccccccccc}
\tabletypesize{\scriptsize}
\tablewidth{0pt}
\tablecaption{Main {\it Chandra} Catalog: Basic Source Properties}

\tablehead{
\colhead{XID} &
\colhead{CXOU Name}                   &
\colhead{RA (J2000)}                   &
\colhead{Dec (J2000)}                   &
\colhead{Dist (') }                   &
\colhead{PU (")}                   &
\colhead{$\log L_{X}$}                   &
\colhead{Var}                   &
\colhead{$\sigma_{\rm var}$}                   &
\colhead{Flag$_{\rm D_{25}}$} &
\colhead{Note}   &
\colhead{GCID}    \\
\colhead{(1)}         &
\colhead{(2)}         &
\colhead{(3)}         &
\colhead{(4)}         &
\colhead{(5)}         &
\colhead{(6)}         &
\colhead{(7)}         &
\colhead{(8)}         &
\colhead{(9)}         &
\colhead{(10)}         &
\colhead{(11)}         &
\colhead{(12)}         \\
}

\startdata
  1&     J124320.4$+$113028&  12:43:20.42&$+$11:30:27.5& 5.50&  1.4&\phantom{$<$}38.28&      N& 0.1&  0&  3&     L282\\
  2&     J124322.6$+$112946&  12:43:22.57&$+$11:29:46.3& 5.45&  1.1&\phantom{$<$}38.06&      V& 0.9&  0&  0&      ...\\
  3&     J124323.1$+$113304&  12:43:23.06&$+$11:33:04.0& 4.14&  1.2&\phantom{$<$}37.53&      N& 0.6&  0&  0&      ...\\
  4&     J124323.2$+$113217&  12:43:23.18&$+$11:32:17.1& 4.21&  2.1&          $<$37.45&      N&-1.0&  0&  0&      ...\\
  5&     J124323.2$+$113037&  12:43:23.25&$+$11:30:37.1& 4.82&  0.6&\phantom{$<$}38.12&      N& 0.7&  0&  3&     L374\\
  6&     J124323.5$+$113109&  12:43:23.47&$+$11:31:09.0& 4.52&  0.9&\phantom{$<$}37.94&      V& 2.2&  0&  3&     L223\\
  7&     J124324.2$+$113109&  12:43:24.21&$+$11:31:09.3& 4.35&  0.9&\phantom{$<$}37.92&      N& 0.0&  0&  0&      ...\\
  8&     J124324.3$+$113428&  12:43:24.29&$+$11:34:28.3& 4.06&  0.7&\phantom{$<$}37.58&      N& 1.2&  0&  0&      ...\\
\enddata

\tablecomments{ 
Col. (1): source number.
Col. (2): IAU name.
Cols. (3) and (4): source right ascension and declination.
Col. (5): radial distance of the source to the nucleus, in units of arcminutes.
Col. (6): positional uncertainty, in units of arcseconds.
Col. (7): logarithmic FB luminosity, in units of erg s$^{-1}$. A 3$\sigma$ upper limit is given if the source is not detected in the FB.
Col. (8): long-term variability flag (see Section 3.4). The source is labeled as variable (V), non-variable (N), transient candidate (TC), or possible transient candidate (PTC).
Col. (9): maximum statistical significance of the FB flux variation between any two observations (see Section 3.4). It is set to ``$-1.0$' if this flux variation is not available.
Col. (10): positional flag, outside the D$_{25}$ ellipses of NGC 4649 and NGC 4647 (``0'), within the D$_{25}$ ellipses of NGC 4649 only (``1'), within the D$_{25}$ ellipses of NGC 4647 only (``2'), or within both D$_{25}$ ellipses (``3').
Col. (11): note on the optical association; see Section~\ref{sec-catalog} for details.
Col. (12): GC ID from J.~Strader et al. (2012, in prep.) or Lee et al. (2008; starting with the letter ``L''); see Section~\ref{sec-catalog} for details. (This table is available in its entirety in a machine-readable form in the online journal. A portion is shown here for guidance regarding its form and content.)}
\label{tbl-main}
\end{deluxetable}

%
%

\begin{deluxetable}{llllllrrrc}
\tabletypesize{\footnotesize}
\tablewidth{0pt}
\tablecaption{Source Counts, Hardness Ratios, Color-Color Values}

\tablehead{
\colhead{}                   &
\multicolumn{5}{c}{Net Counts} &
\colhead{}                   &
\colhead{}                   &
\colhead{}                   \\
\\ \cline{2-6} \\
\colhead{XID}                   &
\colhead{FB}                   &
\colhead{SB}                   &
\colhead{HB}                   &
\colhead{SB1}                   &
\colhead{SB2}                   &
\colhead{HR}                   &
\colhead{SC}                   &
\colhead{HC}                   &
\colhead{$\log L_{X}$}         \\
\colhead{(1)}         &
\colhead{(2)}         &
\colhead{(3)}         &
\colhead{(4)}         &
\colhead{(5)}         &
\colhead{(6)}         &
\colhead{(7)}         &
\colhead{(8)}         &
\colhead{(9)}         &
\colhead{(10)}         \\
}

\startdata
  1&\phantom{$<$}$  23.6  _{-5.5}^{+6.7}$&\phantom{$<$}$  17.2  _{-4.5}^{+5.7}$&\phantom{$<$}$   6.5  _{-3.2}^{+4.6}$&          $<$$  13.2                $&\phantom{$<$}$  14.7  _{-4.1}^{+5.3}$&$ -0.58_{-0.21}^{+0.17}$&$  0.37_{-0.25}^{+0.22}$&$ -0.34_{-0.21}^{+0.26}$&\phantom{$<$}38.28\\
  2&\phantom{$<$}$  33.5  _{-7.1}^{+8.3}$&\phantom{$<$}$  24.9  _{-5.6}^{+6.8}$&          $<$$  26.9                $&\phantom{$<$}$  11.8  _{-3.7}^{+5.0}$&\phantom{$<$}$  13.2  _{-4.1}^{+5.4}$&$ -0.60_{-0.19}^{+0.19}$&$ -0.16_{-0.21}^{+0.27}$&$ -0.15_{-0.17}^{+0.19}$&\phantom{$<$}38.06\\
  3&\phantom{$<$}$  15.9  _{-5.1}^{+6.4}$&\phantom{$<$}$   8.3  _{-3.6}^{+4.9}$&\phantom{$<$}$   7.9  _{-3.7}^{+5.1}$&          $<$$   8.8                $&\phantom{$<$}$   9.2  _{-3.5}^{+4.8}$&$ -0.20_{-0.35}^{+0.29}$&$  0.47_{-0.34}^{+0.20}$&$ -0.12_{-0.34}^{+0.37}$&\phantom{$<$}37.53\\
  4&          $<$$  18.0                $&          $<$$  14.5                $&          $<$$  14.0                $&          $<$$  10.5                $&          $<$$  12.7                $&$ -0.29_{-0.71}^{+0.09}$&$  0.82_{-1.82}^{+0.18}$&$ -0.59_{-0.41}^{+1.59}$&          $<$37.45\\
  5&\phantom{$<$}$  75.5 _{-9.8}^{+11.0}$&\phantom{$<$}$  59.0  _{-8.3}^{+9.5}$&\phantom{$<$}$  16.7  _{-5.2}^{+6.5}$&\phantom{$<$}$  16.4  _{-4.5}^{+5.8}$&\phantom{$<$}$  42.9  _{-7.0}^{+8.3}$&$ -0.65_{-0.10}^{+0.09}$&$  0.18_{-0.14}^{+0.14}$&$ -0.33_{-0.11}^{+0.13}$&\phantom{$<$}38.12\\
  6&\phantom{$<$}$  31.7  _{-6.8}^{+8.0}$&\phantom{$<$}$  20.5  _{-5.1}^{+6.4}$&\phantom{$<$}$  11.6  _{-4.6}^{+5.9}$&\phantom{$<$}$   4.9  _{-2.6}^{+4.0}$&\phantom{$<$}$  15.7  _{-4.4}^{+5.7}$&$ -0.42_{-0.21}^{+0.19}$&$  0.20_{-0.21}^{+0.22}$&$ -0.15_{-0.19}^{+0.22}$&\phantom{$<$}37.94\\
  7&\phantom{$<$}$  31.2  _{-6.7}^{+7.9}$&\phantom{$<$}$  18.0  _{-4.9}^{+6.1}$&\phantom{$<$}$  13.7  _{-4.7}^{+6.1}$&          $<$$  15.6                $&\phantom{$<$}$  14.7  _{-4.3}^{+5.5}$&$ -0.27_{-0.22}^{+0.19}$&$  0.27_{-0.20}^{+0.19}$&$ -0.05_{-0.22}^{+0.22}$&\phantom{$<$}37.92\\
  8&\phantom{$<$}$  41.9  _{-7.7}^{+8.9}$&\phantom{$<$}$  28.8  _{-6.1}^{+7.3}$&\phantom{$<$}$  13.4  _{-4.8}^{+6.1}$&          $<$$  16.8                $&\phantom{$<$}$  25.1  _{-5.5}^{+6.8}$&$ -0.49_{-0.17}^{+0.14}$&$  0.39_{-0.19}^{+0.17}$&$ -0.27_{-0.17}^{+0.20}$&\phantom{$<$}37.58\\
\enddata

\tablecomments{ 
Col. (1): source number.
Cols. (2)--(6): source counts and the associated 1$\sigma$ errors in the FB, SB, HB, SB1, and SB2. A 3$\sigma$ upper limit is given if the source is not detected in the given band.
Col. (7): hardness ratio and the associated 1$\sigma$ errors.
Cols. (8) and (9): soft and hard X-ray colors and the associated 1$\sigma$ errors.
Col. (10): logarithmic FB luminosity, in units of erg s$^{-1}$. A 3$\sigma$ upper limit is given if the source is not detected in the FB. (This table is available in its entirety in a machine-readable form in the online journal. A portion is shown here for guidance regarding its form and content.)}
\label{tbl-mainadd}

\end{deluxetable}

%
%

\begin{deluxetable}{lrrrrrrrrc}
\tabletypesize{\footnotesize}
\tablewidth{0pt}
\tablecaption{Source Counts, Hardness Ratios, Color Values: Observation 1}

\tablehead{
\colhead{}                   &
\multicolumn{5}{c}{Net Counts} &
\colhead{}                   &
\colhead{}                   &
\colhead{}                   \\
\\ \cline{2-6} \\
\colhead{XID}                   &
\colhead{FB}                   &
\colhead{SB}                   &
\colhead{HB}                   &
\colhead{SB1}                   &
\colhead{SB2}                   &
\colhead{HR}                   &
\colhead{SC}                   &
\colhead{HC}                   &
\colhead{$\log L_{X}$}         \\
\colhead{(1)}         &
\colhead{(2)}         &
\colhead{(3)}         &
\colhead{(4)}         &
\colhead{(5)}         &
\colhead{(6)}         &
\colhead{(7)}         &
\colhead{(8)}         &
\colhead{(9)}         &
\colhead{(10)}         \\
}

\startdata
  1& ...& ...& ...& ...& ...& ...& ...& ...& ...\\
  2& ...& ...& ...& ...& ...& ...& ...& ...& ...\\
  3& ...& ...& ...& ...& ...& ...& ...& ...& ...\\
  4& ...& ...& ...& ...& ...& ...& ...& ...& ...\\
  5& ...& ...& ...& ...& ...& ...& ...& ...& ...\\
  6& ...& ...& ...& ...& ...& ...& ...& ...& ...\\
  7& ...& ...& ...& ...& ...& ...& ...& ...& ...\\
  8&\phantom{$<$}$   7.8  _{-3.1}^{+4.4}$&\phantom{$<$}$   4.8  _{-2.4}^{+3.7}$&          $<$$  14.6                $&          $<$$   9.4                $&\phantom{$<$}$   4.1  _{-2.1}^{+3.5}$&$   -0.26_{-0.43}^{+0.32}$&$    0.37_{-0.44}^{+0.31}$&$   -0.12_{-0.43}^{+0.48}$&\phantom{$<$}37.73\\
\enddata

\tablecomments{Columns are the same as those in Table~\ref{tbl-mainadd} but for observation 1. There are no entries for sources not covered by this observation. (This table is available in its entirety in a machine-readable form in the online journal. A portion is shown here for guidance regarding its form and content.)} 
\label{tbl-ind1}

\end{deluxetable}

%
%

\begin{deluxetable}{lrrrrrrrrc}
\tabletypesize{\footnotesize}
\tablewidth{0pt}
\tablecaption{Source Counts, Hardness Ratios, Color Values: Observation 2}

\tablehead{
\colhead{}                   &
\multicolumn{5}{c}{Net Counts} &
\colhead{}                   &
\colhead{}                   &
\colhead{}                   \\
\\ \cline{2-6} \\
\colhead{XID}                   &
\colhead{FB}                   &
\colhead{SB}                   &
\colhead{HB}                   &
\colhead{SB1}                   &
\colhead{SB2}                   &
\colhead{HR}                   &
\colhead{SC}                   &
\colhead{HC}                   &
\colhead{$\log L_{X}$}         \\
\colhead{(1)}         &
\colhead{(2)}         &
\colhead{(3)}         &
\colhead{(4)}         &
\colhead{(5)}         &
\colhead{(6)}         &
\colhead{(7)}         &
\colhead{(8)}         &
\colhead{(9)}         &
\colhead{(10)}         \\
}

\startdata
  1&\phantom{$<$}$  18.1  _{-4.8}^{+6.0}$&\phantom{$<$}$  12.1  _{-3.7}^{+5.0}$&\phantom{$<$}$   6.1  _{-3.0}^{+4.4}$&          $<$$  11.4                $&\phantom{$<$}$  10.6  _{-3.4}^{+4.7}$&$   -0.47_{-0.25}^{+0.20}$&$    0.36_{-0.27}^{+0.24}$&$   -0.25_{-0.24}^{+0.30}$&\phantom{$<$}38.29\\
  2&\phantom{$<$}$  21.4  _{-5.5}^{+6.7}$&\phantom{$<$}$  19.2  _{-4.8}^{+6.0}$&          $<$$  15.4                $&\phantom{$<$}$   9.3  _{-3.2}^{+4.5}$&\phantom{$<$}$   9.9  _{-3.5}^{+4.8}$&$   -0.84_{-0.16}^{+0.05}$&$   -0.17_{-0.28}^{+0.35}$&$   -0.34_{-0.16}^{+0.25}$&\phantom{$<$}38.13\\
  3&          $<$$  10.8                $&          $<$$  11.5                $&          $<$$   7.9                $&          $<$$   7.2                $&          $<$$  11.8                $&$   -0.62_{-0.38}^{+0.01}$&$    0.32_{-1.32}^{+0.68}$&$   -0.99_{-0.01}^{+1.99}$&          $<$38.00\\
  4&          $<$$  10.6                $&          $<$$   9.2                $&          $<$$   9.8                $&          $<$$   7.2                $&          $<$$   9.7                $&$   -0.10_{-0.90}^{+0.18}$&$    0.86_{-1.86}^{+0.14}$&$   -0.78_{-0.22}^{+1.78}$&          $<$37.85\\
  5&\phantom{$<$}$  26.2  _{-5.7}^{+6.9}$&\phantom{$<$}$  20.8  _{-4.9}^{+6.1}$&\phantom{$<$}$   5.4  _{-2.8}^{+4.3}$&\phantom{$<$}$   4.9  _{-2.4}^{+3.7}$&\phantom{$<$}$  16.1  _{-4.3}^{+5.5}$&$   -0.66_{-0.19}^{+0.14}$&$    0.32_{-0.26}^{+0.22}$&$   -0.40_{-0.17}^{+0.26}$&\phantom{$<$}38.21\\
  6&\phantom{$<$}$  18.3  _{-4.8}^{+6.1}$&\phantom{$<$}$   9.5  _{-3.4}^{+4.7}$&\phantom{$<$}$   9.2  _{-3.5}^{+4.9}$&          $<$$  13.0                $&\phantom{$<$}$   7.2  _{-2.9}^{+4.2}$&$   -0.12_{-0.28}^{+0.24}$&$    0.20_{-0.24}^{+0.22}$&$    0.08_{-0.28}^{+0.27}$&\phantom{$<$}38.06\\
  7&\phantom{$<$}$  10.2  _{-3.6}^{+4.9}$&\phantom{$<$}$   5.4  _{-2.6}^{+4.0}$&\phantom{$<$}$   4.9  _{-2.6}^{+4.0}$&          $<$$  11.1                $&\phantom{$<$}$   4.2  _{-2.1}^{+3.5}$&$   -0.13_{-0.39}^{+0.33}$&$    0.22_{-0.34}^{+0.32}$&$    0.05_{-0.40}^{+0.39}$&\phantom{$<$}37.79\\
  8&          $<$$  11.3                $&          $<$$  11.6                $&          $<$$   7.9                $&          $<$$   7.2                $&          $<$$  11.9                $&$   -0.65_{-0.35}^{+0.01}$&$    0.75_{-1.75}^{+0.25}$&$   -0.77_{-0.23}^{+1.77}$&          $<$38.07\\
\enddata

\tablecomments{Columns are the same as those in Table~\ref{tbl-mainadd} but for observation 2. There are no entries for sources not covered by this observation. (This table is available in its entirety in a machine-readable form in the online journal. A portion is shown here for guidance regarding its form and content.)} 
\label{tbl-ind2}

\end{deluxetable}

%
%

\begin{deluxetable}{lrrrrrrrrc}
\tabletypesize{\footnotesize}
\tablewidth{0pt}
\tablecaption{Source Counts, Hardness Ratios, Color Values: Observation 3}

\tablehead{
\colhead{}                   &
\multicolumn{5}{c}{Net Counts} &
\colhead{}                   &
\colhead{}                   &
\colhead{}                   \\
\\ \cline{2-6} \\
\colhead{XID}                   &
\colhead{FB}                   &
\colhead{SB}                   &
\colhead{HB}                   &
\colhead{SB1}                   &
\colhead{SB2}                   &
\colhead{HR}                   &
\colhead{SC}                   &
\colhead{HC}                   &
\colhead{$\log L_{X}$}         \\
\colhead{(1)}         &
\colhead{(2)}         &
\colhead{(3)}         &
\colhead{(4)}         &
\colhead{(5)}         &
\colhead{(6)}         &
\colhead{(7)}         &
\colhead{(8)}         &
\colhead{(9)}         &
\colhead{(10)}         \\
}

\startdata
  1&\phantom{$<$}$   5.6  _{-2.6}^{+4.0}$&\phantom{$<$}$   5.1  _{-2.4}^{+3.7}$&          $<$$   9.9                $&          $<$$   9.6                $&\phantom{$<$}$   4.2  _{-2.1}^{+3.5}$&$   -0.79_{-0.21}^{+0.04}$&$    0.36_{-0.65}^{+0.58}$&$   -0.61_{-0.34}^{+0.63}$&\phantom{$<$}38.25\\
  2&\phantom{$<$}$  12.0  _{-3.9}^{+5.2}$&\phantom{$<$}$   4.6  _{-2.4}^{+3.7}$&\phantom{$<$}$   7.7  _{-3.2}^{+4.6}$&          $<$$  11.4                $&          $<$$  13.6                $&$    0.15_{-0.32}^{+0.34}$&$    0.03_{-0.24}^{+0.26}$&$    0.38_{-0.42}^{+0.30}$&\phantom{$<$}38.33\\
  3&          $<$$   9.5                $&          $<$$   7.3                $&          $<$$  10.4                $&          $<$$   7.2                $&          $<$$   7.4                $&$    0.30_{-0.10}^{+0.70}$&$   -0.09_{-0.84}^{+0.85}$&$    0.83_{-1.83}^{+0.17}$&          $<$38.41\\
  4&          $<$$   7.4                $&          $<$$   7.3                $&          $<$$   7.8                $&          $<$$   7.2                $&          $<$$   7.4                $&$   -0.07_{-0.93}^{+0.40}$&$   -0.00_{-1.00}^{+1.00}$&$    0.06_{-1.06}^{+0.94}$&          $<$38.14\\
  5&\phantom{$<$}$  12.0  _{-3.8}^{+5.1}$&\phantom{$<$}$   9.3  _{-3.2}^{+4.6}$&          $<$$  14.3                $&\phantom{$<$}$   5.2  _{-2.4}^{+3.7}$&\phantom{$<$}$   4.2  _{-2.1}^{+3.5}$&$   -0.62_{-0.29}^{+0.16}$&$   -0.28_{-0.32}^{+0.47}$&$   -0.13_{-0.25}^{+0.28}$&\phantom{$<$}38.33\\
  6&\phantom{$<$}$   4.4  _{-2.4}^{+3.8}$&\phantom{$<$}$   5.0  _{-2.4}^{+3.7}$&          $<$$   7.9                $&          $<$$  11.8                $&          $<$$  13.7                $&$   -0.87_{-0.13}^{+0.01}$&$   -0.06_{-0.80}^{+0.81}$&$   -0.59_{-0.33}^{+0.68}$&\phantom{$<$}37.90\\
  7&          $<$$  13.0                $&          $<$$  11.5                $&          $<$$  10.0                $&          $<$$   9.7                $&          $<$$   9.5                $&$   -0.36_{-0.64}^{+0.15}$&$   -0.39_{-0.61}^{+1.12}$&$   -0.00_{-0.78}^{+0.83}$&          $<$38.35\\
  8&          $<$$  11.6                $&          $<$$   7.3                $&          $<$$  12.5                $&          $<$$   7.2                $&          $<$$   7.4                $&$    0.50_{-0.08}^{+0.50}$&$    0.74_{-1.74}^{+0.26}$&$   -0.98_{-0.02}^{+1.98}$&          $<$38.55\\
\enddata

\tablecomments{Columns are the same as those in Table~\ref{tbl-mainadd} but for observation 3. There are no entries for sources not covered by this observation. (This table is available in its entirety in a machine-readable form in the online journal. A portion is shown here for guidance regarding its form and content.)} 
\label{tbl-ind3}

\end{deluxetable}

%
%

\begin{deluxetable}{lrrrrrrrrc}
\tabletypesize{\footnotesize}
\tablewidth{0pt}
\tablecaption{Source Counts, Hardness Ratios, Color Values: Observation 4}

\tablehead{
\colhead{}                   &
\multicolumn{5}{c}{Net Counts} &
\colhead{}                   &
\colhead{}                   &
\colhead{}                   \\
\\ \cline{2-6} \\
\colhead{XID}                   &
\colhead{FB}                   &
\colhead{SB}                   &
\colhead{HB}                   &
\colhead{SB1}                   &
\colhead{SB2}                   &
\colhead{HR}                   &
\colhead{SC}                   &
\colhead{HC}                   &
\colhead{$\log L_{X}$}         \\
\colhead{(1)}         &
\colhead{(2)}         &
\colhead{(3)}         &
\colhead{(4)}         &
\colhead{(5)}         &
\colhead{(6)}         &
\colhead{(7)}         &
\colhead{(8)}         &
\colhead{(9)}         &
\colhead{(10)}         \\
}

\startdata
  1& ...& ...& ...& ...& ...& ...& ...& ...& ...\\
  2&          $<$$  11.5                $&          $<$$  11.0                $&          $<$$   9.5                $&          $<$$   9.5                $&          $<$$   9.3                $&$   -0.40_{-0.60}^{+0.03}$&$    0.71_{-1.71}^{+0.29}$&$    0.20_{-1.20}^{+0.80}$&          $<$38.21\\
  3&\phantom{$<$}$   6.9  _{-3.3}^{+4.6}$&          $<$$  15.8                $&          $<$$  15.9                $&          $<$$   9.1                $&          $<$$  15.2                $&$   -0.16_{-0.59}^{+0.34}$&$    0.30_{-0.50}^{+0.43}$&$   -0.07_{-0.52}^{+0.55}$&\phantom{$<$}37.59\\
  4&          $<$$   7.4                $&          $<$$   7.3                $&          $<$$   7.9                $&          $<$$   7.3                $&          $<$$   7.4                $&$   -0.09_{-0.91}^{+0.36}$&$    0.70_{-1.70}^{+0.30}$&$    0.11_{-1.11}^{+0.89}$&          $<$37.58\\
  5&\phantom{$<$}$  37.3  _{-6.9}^{+8.2}$&\phantom{$<$}$  28.8  _{-5.8}^{+7.1}$&\phantom{$<$}$   8.6  _{-3.7}^{+5.1}$&\phantom{$<$}$   6.4  _{-2.9}^{+4.2}$&\phantom{$<$}$  22.7  _{-5.1}^{+6.4}$&$   -0.66_{-0.15}^{+0.12}$&$    0.22_{-0.21}^{+0.21}$&$   -0.35_{-0.15}^{+0.20}$&\phantom{$<$}38.21\\
  6&\phantom{$<$}$   8.9  _{-4.0}^{+5.3}$&\phantom{$<$}$   6.0  _{-2.9}^{+4.2}$&          $<$$  17.0                $&          $<$$   9.3                $&\phantom{$<$}$   5.6  _{-2.7}^{+4.0}$&$   -0.52_{-0.27}^{+0.15}$&$    0.43_{-0.52}^{+0.40}$&$   -0.31_{-0.45}^{+0.54}$&\phantom{$<$}37.51\\
  7&\phantom{$<$}$  18.9  _{-5.2}^{+6.5}$&\phantom{$<$}$  10.9  _{-3.8}^{+5.0}$&\phantom{$<$}$   8.3  _{-3.7}^{+5.1}$&          $<$$  11.0                $&\phantom{$<$}$   9.9  _{-3.5}^{+4.8}$&$   -0.30_{-0.28}^{+0.25}$&$    0.36_{-0.27}^{+0.21}$&$   -0.12_{-0.28}^{+0.31}$&\phantom{$<$}37.82\\
  8&\phantom{$<$}$  13.2  _{-4.3}^{+5.6}$&\phantom{$<$}$   7.0  _{-3.0}^{+4.4}$&\phantom{$<$}$   6.4  _{-3.1}^{+4.5}$&          $<$$  13.3                $&\phantom{$<$}$   4.4  _{-2.4}^{+3.8}$&$   -0.19_{-0.35}^{+0.29}$&$   -0.05_{-0.29}^{+0.36}$&$    0.12_{-0.34}^{+0.30}$&\phantom{$<$}37.66\\
\enddata

\tablecomments{Columns are the same as those in Table~\ref{tbl-mainadd} but for observation 4. There are no entries for sources not covered by this observation. (This table is available in its entirety in a machine-readable form in the online journal. A portion is shown here for guidance regarding its form and content.)} 
\label{tbl-ind4}
\end{deluxetable}

%
%

\begin{deluxetable}{lrrrrrrrrc}
\tabletypesize{\footnotesize}
\tablewidth{0pt}
\tablecaption{Source Counts, Hardness Ratios, Color Values: Observation 5}

\tablehead{
\colhead{}                   &
\multicolumn{5}{c}{Net Counts} &
\colhead{}                   &
\colhead{}                   &
\colhead{}                   \\
\\ \cline{2-6} \\
\colhead{XID}                   &
\colhead{FB}                   &
\colhead{SB}                   &
\colhead{HB}                   &
\colhead{SB1}                   &
\colhead{SB2}                   &
\colhead{HR}                   &
\colhead{SC}                   &
\colhead{HC}                   &
\colhead{$\log L_{X}$}         \\
\colhead{(1)}         &
\colhead{(2)}         &
\colhead{(3)}         &
\colhead{(4)}         &
\colhead{(5)}         &
\colhead{(6)}         &
\colhead{(7)}         &
\colhead{(8)}         &
\colhead{(9)}         &
\colhead{(10)}         \\
}

\startdata
  1& ...& ...& ...& ...& ...& ...& ...& ...& ...\\
  2& ...& ...& ...& ...& ...& ...& ...& ...& ...\\
  3&\phantom{$<$}$   7.1  _{-3.1}^{+4.5}$&          $<$$  13.2                $&\phantom{$<$}$   5.0  _{-2.6}^{+4.1}$&          $<$$   7.3                $&          $<$$  13.8                $&$    0.22_{-0.45}^{+0.44}$&$    0.30_{-0.37}^{+0.28}$&$    0.24_{-0.58}^{+0.47}$&\phantom{$<$}37.80\\
  4&\phantom{$<$}$   6.0  _{-3.2}^{+4.5}$&          $<$$  14.9                $&          $<$$  15.4                $&          $<$$  11.6                $&          $<$$  11.6                $&$   -0.23_{-0.56}^{+0.31}$&$   -0.25_{-0.49}^{+0.64}$&$    0.17_{-0.53}^{+0.45}$&\phantom{$<$}37.52\\
  5& ...& ...& ...& ...& ...& ...& ...& ...& ...\\
  6& ...& ...& ...& ...& ...& ...& ...& ...& ...\\
  7& ...& ...& ...& ...& ...& ...& ...& ...& ...\\
  8&\phantom{$<$}$  18.3  _{-5.0}^{+6.3}$&\phantom{$<$}$  15.5  _{-4.4}^{+5.7}$&          $<$$  15.2                $&          $<$$  10.9                $&\phantom{$<$}$  14.7  _{-4.1}^{+5.4}$&$   -0.79_{-0.11}^{+0.06}$&$    0.61_{-0.36}^{+0.24}$&$   -0.61_{-0.22}^{+0.37}$&\phantom{$<$}37.91\\
\enddata

\tablecomments{Columns are the same as those in Table~\ref{tbl-mainadd} but for observation 5. There are no entries for sources not covered by this observation. (This table is available in its entirety in a machine-readable form in the online journal. A portion is shown here for guidance regarding its form and content.)} 
\label{tbl-ind5}

\end{deluxetable}

%
%

\begin{deluxetable}{lrrrrrrrrc}
\tabletypesize{\footnotesize}
\tablewidth{0pt}
\tablecaption{Source Counts, Hardness Ratios, Color Values: Observation 6}

\tablehead{
\colhead{}                   &
\multicolumn{5}{c}{Net Counts} &
\colhead{}                   &
\colhead{}                   &
\colhead{}                   \\
\\ \cline{2-6} \\
\colhead{XID}                   &
\colhead{FB}                   &
\colhead{SB}                   &
\colhead{HB}                   &
\colhead{SB1}                   &
\colhead{SB2}                   &
\colhead{HR}                   &
\colhead{SC}                   &
\colhead{HC}                   &
\colhead{$\log L_{X}$}         \\
\colhead{(1)}         &
\colhead{(2)}         &
\colhead{(3)}         &
\colhead{(4)}         &
\colhead{(5)}         &
\colhead{(6)}         &
\colhead{(7)}         &
\colhead{(8)}         &
\colhead{(9)}         &
\colhead{(10)}         \\
}

\startdata
  1& ...& ...& ...& ...& ...& ...& ...& ...& ...\\
  2& ...& ...& ...& ...& ...& ...& ...& ...& ...\\
  3&          $<$$   9.9                $&          $<$$   9.8                $&          $<$$   8.1                $&          $<$$   7.3                $&          $<$$  10.0                $&$   -0.54_{-0.46}^{+0.06}$&$    0.60_{-1.60}^{+0.40}$&$   -0.66_{-0.34}^{+1.66}$&          $<$38.75\\
  4&          $<$$   7.5                $&          $<$$   7.3                $&          $<$$   8.0                $&          $<$$   7.3                $&          $<$$   7.4                $&$   -0.07_{-0.93}^{+0.40}$&$    0.01_{-1.01}^{+0.99}$&$    0.06_{-1.06}^{+0.94}$&          $<$38.40\\
  5& ...& ...& ...& ...& ...& ...& ...& ...& ...\\
  6& ...& ...& ...& ...& ...& ...& ...& ...& ...\\
  7& ...& ...& ...& ...& ...& ...& ...& ...& ...\\
  8&          $<$$   7.5                $&          $<$$   7.3                $&          $<$$   7.9                $&          $<$$   7.3                $&          $<$$   7.4                $&$   -0.07_{-0.93}^{+0.39}$&$    0.01_{-1.01}^{+0.99}$&$    0.06_{-1.06}^{+0.94}$&          $<$38.34\\
\enddata

\tablecomments{Columns are the same as those in Table~\ref{tbl-mainadd} but for observation 6. There are no entries for sources not covered by this observation. (This table is available in its entirety in a machine-readable form in the online journal. A portion is shown here for guidance regarding its form and content.)} 
\label{tbl-ind6}

\end{deluxetable}

\end{document}